%% file: main.tex
\documentclass{aa}
\usepackage[varg]{txfonts}
\usepackage{graphicx}
\usepackage{natbib}
\bibpunct{(}{)}{;}{a}{}{,} 

\begin{document}

\title{Characterisation of chaos and mean-motion resonances in meteoroid streams}
\subtitle{Application to the Draconids and Leonids}
\author{Ariane Courtot \thanks{ariane.courtot@obspm.fr} \and Melaine Saillenfest \and Jérémie Vaubaillon \and Marc Fouchard}
\institute{IMCCE, Observatoire de Paris, PSL Research University, CNRS, Sorbonne Universit\'{e}, UPMC Univ Paris 6, Univ Lille, France}
\date{Received May 28, 2023 / Accepted date }

\abstract
{Dynamically linking a meteor shower with its parent body is challenging, and chaos in the dynamics of meteoroid streams may be one of the reasons. For a robust identification of parent bodies, it is therefore necessary to quantify the amount of chaos involved in the evolution of meteoroid streams.}
{The characterisation of chaos in meteoroid streams thanks to chaos maps is still a new field of study. We aim to study two very different meteoroid streams, the Draconids and the Leonids, in order to obtain a general view of this topic.}
{We use the method developed in a previous paper dedicated to Geminids, drawing chaos maps with the orthogonal fast Lyapunov indicator. We choose four particle size ranges to investigate the effect of non-gravitational forces. As the dynamics is structured by mean-motion resonances with planets, we compute the locations and widths of the resonances at play. We use semi-analytical formulas valid for any eccentricity and inclination and an arbitrary number of planets.}
{We pinpoint which mean-motion resonances with Jupiter play a major role in the dynamics of each meteoroid stream. We show how those resonances tend to trap mostly large particles, preventing them from meeting with Jupiter. We also study particles managing to escape those resonances, thanks to the gravitational perturbation of Saturn for example. Finally, we explain why non-gravitational forces do not disturb the dynamics much, contrary to what is observed for the Geminids.}
{}


\maketitle

\section{Introduction}

The prediction of meteor showers is based on the knowledge of their parent body.
In absence of such knowledge, an hypothesis about their origin must be made first \citep{LyytinenJenniskens2003,VaubaillonJenniskens2007,Jenniskens2008,Jenniskens_2008}.
However, even the initial conditions of the meteoroid stream are determined on the basis of the existence of a meteor shower.
Out of the numerous meteors recorded for the past two decades at least, nearly 1600 showers in total are in the IAU list of meteor showers\footnote{\url{https://www.ta3.sk/IAUC22DB/MDC2022/Roje/roje\_lista.php} accessed in April 2023}.
However, many of them are probably not genuine meteor showers. The very identification of meteor showers is an area of research in itself, and the definition of a robust method to identify them is currently under study \citep{GartrellElford1975,NeslusanHajdukova2017,Jopek2023}.
In particular, a meteor shower can only be recognised if the meteoroid orbits are similar to each other at the time of observation.
This implies that these orbits are also stable enough between the time of their ejection from the parent body and the observation, such that their trajectories did not diverge substantially.
Quantifying the orbit similarity has also been the subject of many efforts in the past, yet yielding inconclusive results \citep{Williams2019}.

In order to help in the effort of establishing the existence of meteor showers, we used a method to quantify the chaoticity in meteoroid streams, as a function of meteoroid size and time \citep{Courtot_al_2023}. Chaoticity is indeed highly size-dependent, as a result of non-gravitational forces, as one could expect \citep{Liou_Zook_1997}.
This first study focused on the Geminids meteoroid stream, for which the parent body asteroid 3200 Phaethon is known to have a very stable orbit for 10 kyrs \citep{WilliamsWu1993,Ryabova2019}.
The Geminids case was therefore a good way to characterise the performances of the tool and its ability to detect chaos in a relatively weak chaotic orbit. Yet, many meteoroid streams are more chaotic than the Geminids. Now that the tools described by \cite{Courtot_al_2023} are well established, we are able to characterise chaos in various streams linked to verified meteor showers, creating a general view of what chaos in meteor showers can look like. This view can serve as a point of reference for future studies of groups of meteors whose identity as meteor showers is not yet certain. Ultimately, it will allow us to distinguish between meteor showers and chance associations of meteoroids.

This paper focuses on two well known, yet very different, meteor showers associated to a Jupiter family comet (JFC) and a Halley-type comet (HTC).
The Draconids meteor shower is caused by the JFC 21P/Giacobini-Zinner, known to have had its last very close encounter with Jupiter in 1898 (closest distance: $0.24$ au).
Such encounters also happen for the associated meteoroid stream, presumably at a much higher frequency, as the stream has expanded a lot in the interplanetary space after a few revolutions, giving it a much larger impact cross section than the comet itself.
Close encounters with giant planets are responsible for the stream dispersion in the Solar System (with the result of gradually loosing the dynamical signature of its parent body because of chaotic diffusion), brutal change of meteoroids orbits and rather complicated processes \citep[for a full review see][]{Vaubaillon2019}.
The Draconids activity is characterised by occasional outbursts of various strength \citep{Egal2019}.
Presumably, the associated meteoroid stream is subject to strong and rapid chaos, that needs to be quantified.

In comparison, the Leonids associated with HTC 55P/Tempel-Tuttle are dynamically less chaotic, since close encounters with Jupiter are less frequent for them.
Their activity is also characterised by dramatic outbursts or even storms \citep{WuWilliams1996,McNaught_Asher99}.
The high level of some shower outbursts is caused by young and densely populated trails colliding with the Earth.
The influence of the Earth in the stream is most of the time negligible and causes small gaps in the trail \citep{Vaubaillon2019}.
In this paper, we aim to quantify the level of chaos in these two drastically different streams.
Following the method developed by \cite{Courtot_al_2023}, chaos maps are drawn as a function of time and meteoroid size.

In Sect. \ref{sec:meth}, the method is briefly recalled and the specificity of the studied streams are described. In Sect.~\ref{sec:methwidth}, a method to compute the width of mean-motion resonances for meteoroids is presented.
In Sect. \ref{sec:results}, the results on the Leonids and Draconids meteoroid streams are shown. More precisely, we explore the role of mean-motion resonances with Jupiter, as well as close encounters with several planets. We also investigate the impact of non-gravitational forces. We conclude in Sect.~\ref{sec:ccl}.

\input{Method}

\input{Results}

\section{Conclusion}\label{sec:ccl}

Amongst the meteor showers listed by the IAU, several are contested: tools available today are not always able to discriminate between actual meteor showers and groups of meteors seemingly coming from similar orbits. In order to contribute to this discussion, we study the influence of chaos on meteoroid streams. We draw chaos maps on well known meteoroid streams that give rise to uncontested meteor showers. In future works, we will compare those maps to others drawn from disputed meteor showers. For this purpose, we need to study different types of meteoroid streams.

After a first study of chaos on the Geminid meteoroid stream, which allowed us to test our method while also yielding interesting results, we tackled this time the JFC Draconid and the HTC Leonid meteoroid streams.
Both their aphelia and perihelia are much greater than the previously studied Geminids, and we therefore expected different dynamics and relationship to chaos compared to the Geminids. 

For both streams, we proved that Jupiter is an important driver of their dynamics. Saturn also plays an important role in the Leonids stream. 

We detected several MMRs with Jupiter and highlighted the phenomenon of capture of particles inside the MMR, preventing them from meeting Jupiter. In the specific case of the Leonids, we showed how close encounters with Saturn inside the Jupiter MMR could sometimes lead to the escape of the particles.

We also studied the effect of non-gravitational forces and noted that this effect is much weaker for the Draconids and Leonids than for the Geminids. In order to explain this, we analysed both the width of the MMRs detected as well as the strength of the non-gravitational forces. We showed that the MMRs involved in the Draconids and the Leonids dynamics are much wider than the MMRs detected for the Geminids. We also showed that the non-gravitational forces are much stronger for the Geminids than for the Draconids or Leonids, due to the shapes of their orbits.

The semi-analytical method presented in Sect.~\ref{sec:methwidth} reveals its full utility when applied to orbits as inclined and eccentric as those of Draconids and Leonids. The detailed structure of mean-motion resonances, including their overall widths, can be computed from a few numerical methods. The agreement between the chaos maps and the semi-analytical results is very encouraging: it shows that the semi-analytical approach can be used in future works not only to characterise resonances, but also to study the dynamical evolution of individual meteoroids as non-gravitational forces make them transition through the inner and outer separatrices of the resonances (either as resonance captures or escape, as small particles gradually leak out of resonance).

From these results, we conclude that, for the streams studied here, the main differences (if any) of dynamical behaviour between large ($>10$ cm) and small ($<100 \: \mu$m) meteoroids mainly comes from their initial ejection velocity from the parent body, but is not related to chaos.

All of this is proven thanks to chaos maps. Maps drawn from other colours are available under request.

In the future, we will dive deeper in the relationship between chaoticity and meteor showers whose existence might be questionable.

\begin{acknowledgements}
Ariane Courtot acknowledges support from the \'{E}cole Doctorale d'Astronomie et d'Astrophysique d'Île-de-France (ED 127).
\end{acknowledgements}

\bibliographystyle{aa} 
\bibliography{biblio.bib}

\appendix
\input{Annexe}

\end{document}

%% file: Method.tex
\section{Method}\label{sec:meth}

We described our method in depth in our previous article \citep{Courtot_al_2023}. We only report here the main points and the specifics of this study.

Drawing chaos maps requires a chaos indicator, and we chose the orthogonal fast Lyapunov indicator (OFLI) from \cite{Fouchard_al_2002} since it suits our problem and purpose. Because of the non-gravitational forces, our problem is dissipative and therefore we cannot use a symplectic integrator. Because we also expect many close encounters, an integrator with a variable time step is preferred. The integrator RADAU order 15 meets this two demands \citep{Everhart_1985}. 

The INPOP planetary ephemerides are used \citep{Fienga_al_2009}. Non-gravitational forces (NGFs) taken into account are the Poynting-Robertson drag and the radiation pressure \citep{Vaubaillon_al_2005}.

Each particle is described by its state vector (position and velocity) at time $t$, and its radius. We assumed a density of $\rho = 1000 \text{kg}\ \text{m}^{-3}$ in order to compute the mass. 

Different initial conditions are chosen for each meteoroid stream. The initial time $t_0$ is set to 1901 AD for the Draconids and 1334 AD for the Leonids. The trail ejected at those times are known to be responsible for meteor outbursts on Earth in 1946 and 1998 respectively \citep{McNaught_Asher99,Vaubaillon_al_2011}.

For each particle, the initial state vector was selected from the ranges of orbital elements described in Table \ref{tab:CIDraco} for the Draconids and Table \ref{tab:CILeo} for the Leonids. 
Those chosen ranges of orbital elements encompass a broad array of orbits characterising those streams, and specifically orbits resulting from simulations using the model developed by \cite{Vaubaillon_al_2005}.
More precisely, for each particle, the orbital elements were picked randomly in those ranges, and the initial state vector computed. We decided on a random selection from our experience with the Geminids \citep[see][]{Courtot_al_2023}. The mean anomaly is not described in the tables because we chose to cover the whole range of possible values (0 to 360°).

We also chose the radius of our particles. We created a first set of particles with a radius randomly chosen between 10 and 100~mm, that we called BIN10100. Then we created three other sets with the same constraints, but with different bins of radii: 1 to 10~mm (BIN110), 0.1 to 1~mm (BIN011) and finally 0.01 to 0.1~mm (BIN00101). We did not consider other sets of smaller particles, because the non-gravitational forces would be heavily modified for such small particles, as the radiation-matter interaction would pass into another regime. The sizes described in the various sets encompass most of what is observed in meteor showers. Each of those sets contains 100080 particles.


\begin{table}
    \caption{Range of heliocentric orbital elements of Draconids}
    \label{tab:CIDraco}
    \centering
    \begin{tabular}{c c c }
    \hline \hline
         Element & Min & Max \\ 
         \hline
         $a$ (au) & 2.8 & 4 \\
         $e$ & 0.66 & 0.81 \\
         $i$ (°) & 28 & 32 \\
         $\omega$ (°) & 168 & 173 \\
         $\Omega$ (°) & 196 & 200 \\ 
         \hline
    \end{tabular}
    \tablefoot{The Draconids are integrated during 1000~years from these ranges of elements.}
\end{table}

\begin{table}
    \caption{Range of heliocentric orbital elements of Leonids}
    \label{tab:CILeo}
    \centering
    \begin{tabular}{c c c}
    \hline \hline
         Element & Min & Max \\
         \hline
         $a$ (au) & 9.8 & 11 \\
         $e$ & 0.83 & 0.98 \\
         $i$ (°) & 165 & 169 \\
         $\omega$ (°) & 226 & 230 \\
         $\Omega$ (°) & 168 & 172 \\ 
         \hline
    \end{tabular}
    \tablefoot{The Leonids are integrated during 2000 years from these ranges of elements.}
\end{table}

We then integrated the Draconids for 1000 years (about 170 periods) and the Leonids for 2000 years (about 60 periods), to study short and mid-term behaviour of the streams, without modelling their entire lifetime.



\section{Computation of resonance widths}\label{sec:methwidth}

   The orbital dynamics of meteoroid streams are shaped by mean-motion resonances with planets. On the chaos maps previously computed by \cite{Courtot_al_2023} for the Geminids, resonances appear as chaotic belts (the resonance `separatrices') surrounding stable zones. Because we expect mean-motion resonances to also play a decisive role in the dynamics of Draconids and Leonids, we need to locate the relevant resonances and determine their widths.
   
   The traditional description of mean-motion resonances relies on the expansion of the disturbing potential in series of eccentricity $e$ and inclination $i$ (see e.g. \citealp{Laskar-Robutel_1995,Murray-Dermott_1999}). For a given resonance, the leading-order term of the averaged series defines the `resonance angle', whose behaviour sets the global dynamics of the system. This classical approach is only valid in the limit of low eccentricities and low mutual inclinations. In the case of meteoroid streams, eccentricities and inclinations can reach any value including regions where the classical series expansion diverges (for instance $e\approx 0.7$ and $i\approx 30^\circ$ for Draconids, and $e\approx 0.9$ and $i\approx 167^\circ$ for Leonids; see Tables~\ref{tab:CIDraco} and \ref{tab:CILeo}). Even though the disturbing potential can in theory be expanded around arbitrary values (see e.g. \citealp{Lei_2019,Namouni-Morais_2020}), this approach still requires to correctly identify the relevant resonant harmonics in the truncated series. This process can be cumbersome for high-eccentricity and high-inclination orbits, since many harmonics simultaneously play a strong role, and their relative importance depends on the varying orbital elements of the particle.
   
   As an alternative to series expansions, one can compute the resonant disturbing potential numerically (see e.g. \citealp{Gomes-etal_2005,Gallardo_2006a,Gallardo_2006b,Gallardo-etal_2012}). In this case, the resulting potential is exact, but the drawback is that we cannot solve the equations of motion explicitly. Yet, a semi-analytical model of the long-term dynamics is still possible by using the so-called `adiabatic invariant' approximation \citep{Lenard_1959,Henrard_1982,Wisdom_1985}. The adiabatic approximation is commonly used in celestial mechanics to reduce a system with two well-separated timescales into a system with fewer degrees of freedom. Of particular interest are systems reducing to only one degree of freedom, as trajectories can be represented by the level curves of a conserved quantity. This is the case of the planar dynamics of small bodies in mean-motion resonance with a planet (see e.g. \citealp{Wisdom_1985,Beust-Morbidelli_1996,Beust_2016,Pichierri-etal_2017}), or the spatial dynamics of small bodies perturbed by planets on coplanar circular orbits (see e.g. \citealp{Gallardo-etal_2012,Saillenfest-etal_2016,Efimov-Sidorenko_2020,Saillenfest_2020,Lei-etal_2022}). Here, we are not interested in obtaining a fully integrable model for the long-term resonant dynamics, but only to compute the resonance widths. Therefore, a large number of degrees of freedom is not an issue as long as their variations are `slow' (see below). Consequently, we may consider arbitrary orbits for both the small body and the planets.
   
   Writing $\varepsilon$ the characteristic size of planetary perturbations, the adiabatic approximation consists in taking advantage of the large separation of timescales between the oscillation of the small body inside a mean-motion resonance (frequency proportional to $\varepsilon^{1/2}$) and the orbital precession of the small body and the planets (frequencies proportional to $\varepsilon^1$). We outline here the semi-analytical method of \cite{Gallardo_2019,Gallardo_2020}, that we adapt to the motion of a small body under the influence of any number of possibly inclined and eccentric planets. The background of the equations is given in Appendix~\ref{asec:widths}; it is shown that this method neglects terms of order $\varepsilon^{3/2}$ (and not $\varepsilon^2$ as it is the case of a standard non-resonant secular theory).
   
   \subsection{Basic equations}
   We consider a small body perturbed by $N$ planets and close to a mean-motion resonance $k_p$:$k$ with a given planet $p$. We write $\mu$ the gravitational parameter of the Sun, and $\lambda$ and $\lambda_p$ the mean longitudes of the small body and of the planet $p$. The resonance widths can be computed by studying the one-degree-of-freedom simplified Hamiltonian function:
   \begin{equation}\label{eq:KW}
      \mathcal{K}(\Sigma,\sigma) = -\frac{1}{2}\alpha(\Sigma-\Sigma_0)^2 + \varepsilon W(\sigma)\,,
   \end{equation}
   where $\Sigma = \sqrt{\mu a}/k$ and $\sigma = k\lambda - k_p\lambda_p$ are conjugate coordinates. We define the resonance centre as $\Sigma_0=\sqrt{\mu a_0}/k$, where the central semi-major axis of the resonance is:
   \begin{equation}\label{eq:a0}
      a_0^{3/2} = \frac{k}{k_p}\frac{\sqrt{\mu}}{n_p}\left(1 + 2\sum_{j\in\mathcal{I}}\frac{\varepsilon\mu_j}{\mu}\right)\,.
   \end{equation}
   Here, $n_p$ is the mean-motion of planet $p$ and $\varepsilon\mu_j$ is the gravitational parameter\footnote{The factor $\varepsilon$ has no explicit definition; it is only used here to highlight the smallness of $\varepsilon\mu_j$ with respect to $\mu$.} of planet $j$. The sum in Eq.~\eqref{eq:a0} is made over the subset $\mathcal{I}$ of planets interior to the orbit of the small body. This small corrector term accounts for the constant shift of the resonance centre due to the presence of other planets in the system (see Appendix~\ref{asec:widths}). The constant factor $\alpha$ in Eq.~\eqref{eq:KW} is equal to:
   \begin{equation}\label{eq:alp}
      \alpha = 3\frac{k^2}{a_0^2}\left(1+2\sum_{j\in\mathcal{I}}\frac{\varepsilon\mu_j}{\mu}\right)\,.
   \end{equation}
   If one would expand $\varepsilon W(\sigma)$ in Eq.~\eqref{eq:KW} using the classical series in eccentricities and inclinations, an infinity of resonance angles featuring $\sigma$ would appear, and they would verify the D'Alembert rules (see e.g. \citealp{Murray-Dermott_1999}). Here, we avoid this kind of expansion and keep all these angles at once in the Hamiltonian function. This requires to compute the function $\varepsilon W(\sigma)$ numerically as:
   \begin{equation}\label{eq:WC}
      \left\{
      \begin{aligned}
         \varepsilon W &= - \sum_{j\neq p}\frac{\varepsilon\mu_j}{4\pi^2}\int_{0}^{2\pi}\!\!\!\int_{0}^{2\pi}\frac{1}{\|\mathbf{r}-\mathbf{r}_j\|}\,\mathrm{d}\lambda\,\mathrm{d}\lambda_j \\
         &\ \ \ - \frac{\varepsilon\mu_p}{2\pi k}\int_{0}^{2\pi k}\left(\frac{1}{\|\mathbf{r}-\mathbf{r}_p\|}-\mathbf{r}\cdot\frac{\mathbf{r}_p}{\|\mathbf{r}_p\|^3}\right)\mathrm{d}\lambda_p\,.
      \end{aligned}
      \right.
   \end{equation}
   In this expression, $\mathbf{r}_j$ is the heliocentric position vector of planet~$j$ and $\mathbf{r}$ is the heliocentric position vector of the small body evaluated at a semi-major axi $a=a_0$. The first integral in Eq.~\eqref{eq:WC} is performed over the mean longitudes $\lambda$ and $\lambda_j$ taken separately. The second integral in Eq.~\eqref{eq:WC} is performed over $\lambda_p$ only, while expressing $\lambda$ as $\lambda = (\sigma + k_p\lambda_p)/k$. Other orbital elements are taken as constants.
   
   As stressed in Appendix~\ref{asec:widths}, the constant orbital elements of the planets to be used in these formulas -- including the mean motion $n_p$ in Eq.~\eqref{eq:a0} -- are secular variables that incorporate the effects of mutual perturbations among all planets. In practice, we can take them from an existing analytical theory (see e.g. \citealp{Bretagnon_1982}, or \citealp{Duriez-Vienne_1991} in the satellite case) or compute them from a preliminary numerical integration of the planetary system (see e.g. \citealp{Lei-etal_2022}). Likewise, the orbital elements $(e,i,\omega,\Omega)$ of the small body must be interpreted as mean variables. Using these as fixed parameters, $\varepsilon W$ can be studied as a mere function of $\sigma$.
   
   One can note that the first term in Eq.~\eqref{eq:WC} does not depend on $\sigma$ and is therefore constant. This is true on the resonant timescale used to define the resonance widths, but not on the secular timescale over which the orbital elements $(e,i,\omega,\Omega)$ of the small body vary. The first term in Eq.~\eqref{eq:WC} must therefore be kept if one wants to build upon this method to perform semi-averaged integrations of the system (see e.g. \citealp{Saillenfest-Lari_2017}) or develop a secular theory.
   
   \subsection{The resonance widths}
   For fixed values of $(e,i,\omega,\Omega)$ and of the planets' mean orbital elements, the system has only one degree of freedom, so it is integrable and any possible trajectory can be represented as a level curve of the Hamiltonian function in the plane $(\Sigma,\sigma)$.
   
\begin{figure}
      \includegraphics[width=\columnwidth]{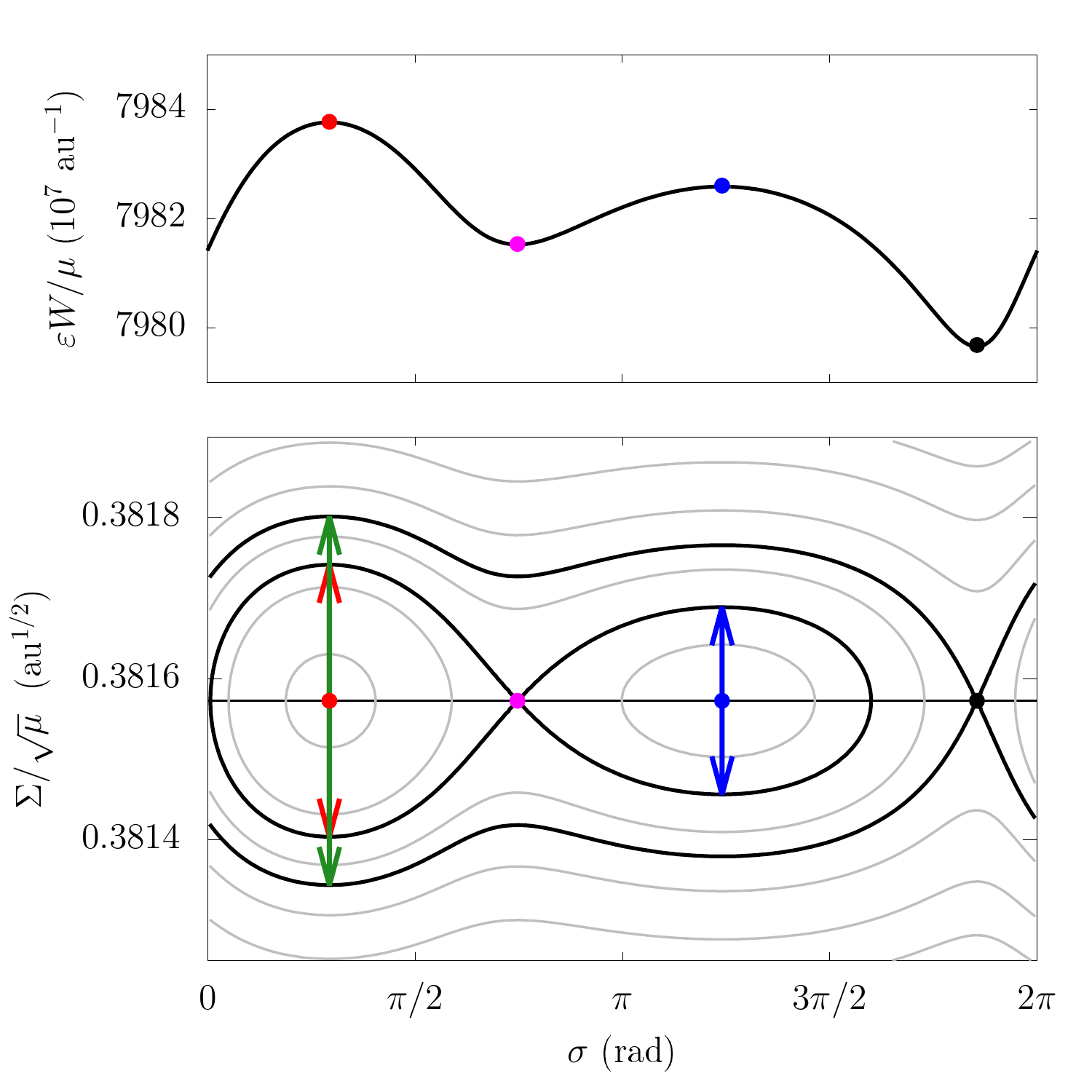}
      \caption{Phase portrait of the resonance $2$:$3$ with the Earth for a small body with $e=0.6$, $i=22^\circ$, $\omega=322^\circ$, and $\Omega=265^\circ$. The bottom panel shows the level curves of the Hamiltonian function in Eq.~\eqref{eq:KW}. The central semi-major axis of the resonance is $a_0=1.31038$~au (horizontal line). Coloured dots show the equilibrium points. The definition of the widths of resonance islands and sub-islands are represented by arrows. The top panel shows the behaviour of $\varepsilon W$ as a function of $\sigma$ (i.e. this is a cut of the bottom panel along the horizontal line).}
      \label{fig:example}
   \end{figure}

   Thanks to the simple form of the Hamiltonian in Eq.~\eqref{eq:KW}, the fixed points can be found easily: all equilibria are located at $\Sigma=\Sigma_0$; stable equilibria are the maxima of $\varepsilon W$ as a function of $\sigma$ and saddle points are its minima. As $\varepsilon W$ is a complex function of $\sigma$ defined by integrals, its maxima and minima as a function of $\sigma$ need to be found numerically (e.g. using Brent's method; see \citealp{Press-etal_2007}). For substantial eccentricities and/or inclinations, many local maxima and minima can be found, resulting in a complex nested structure of the resonance island, as illustrated in Fig.~\ref{fig:example}. This picture strongly differs from the low-eccentricity low-inclination paradigm (see e.g. \citealp{Murray-Dermott_1999}) for which a single cosine term dominates the dynamics with a single critical argument for the resonance.
   
   Once the maxima and minima of $\varepsilon W$ have been found numerically, the Hamiltonian function in Eq.~\eqref{eq:KW} directly gives the widths of all islands and sub-islands of resonance. In particular, the outermost separatrix determines the overall width of the $k_p$:$k$ resonance. If we write $\varepsilon W_\text{max}$ the global maximum of $\varepsilon W$ as a function of $\sigma$ and $\varepsilon W_\text{min}$ its global minimum (red and black dots in Fig.~\ref{fig:example}), the overall half-width of the resonance is:
   \begin{equation}\label{eq:width}
      \Delta\Sigma = \sqrt{\frac{2}{\alpha}(\varepsilon W_\text{max} - \varepsilon W_\text{min})}\,,
   \end{equation}
   illustrated by the green arrow in Fig.~\ref{fig:example}. In terms of the semi-major axis $a$, the corresponding upper and lower boundaries of the resonance are $a = (\sqrt{\mu a_0}\pm k\Delta\Sigma)^2/\mu$. The same formula can be applied on the local minima and maxima of $\varepsilon W$ in order to compute the widths of the sub-islands of resonance.
   
\begin{figure}
      \centering
      \includegraphics[width=0.85\columnwidth]{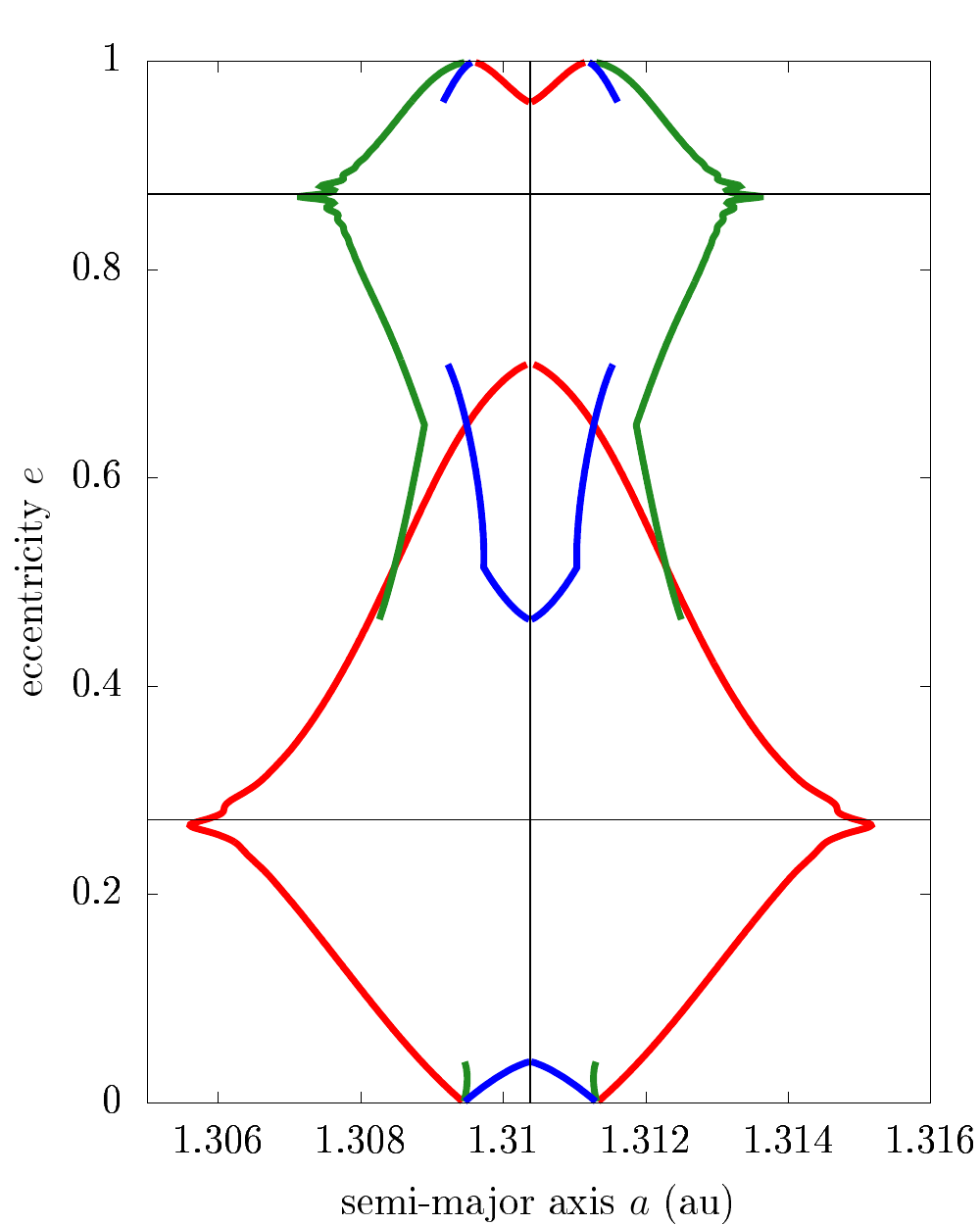}
      \caption{Width of all resonance islands and sub-islands as a function of eccentricity for the $2$:$3$ commensurability with the Earth. The small body has $i=22^\circ$, $\omega=322^\circ$, and $\Omega=265^\circ$ similarly to the Geminids (see \citealp{Courtot_al_2023}). Horizontal lines show the locations where the orbit of the small body crosses the orbit of the resonant planet. See Fig.~\ref{fig:example} for an illustration of the different islands at $e=0.6$.}
      \label{fig:Geminids}
   \end{figure}

\begin{figure}
      \centering
      \includegraphics[width=0.85\columnwidth]{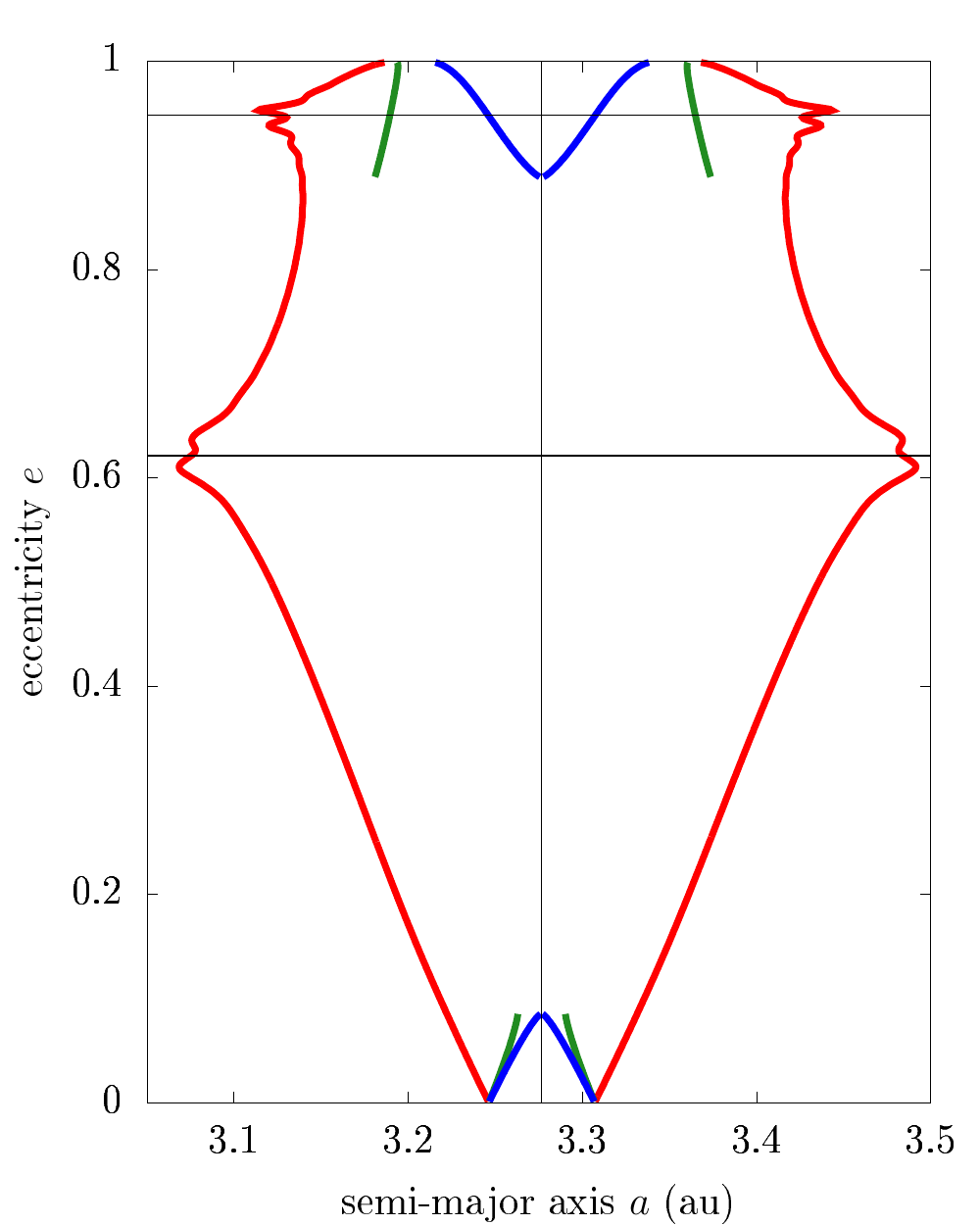}
      \caption{Same as Fig.~\ref{fig:Geminids} for the $2$:$1$ commensurability with Jupiter. The small body has $i=30^\circ$, $\omega=171^\circ$, and $\Omega=198^\circ$, similarly to the Draconids (see Table~\ref{tab:CIDraco}).}
      \label{fig:Draconids}
   \end{figure}

\begin{figure}
      \centering
      \includegraphics[width=0.85\columnwidth]{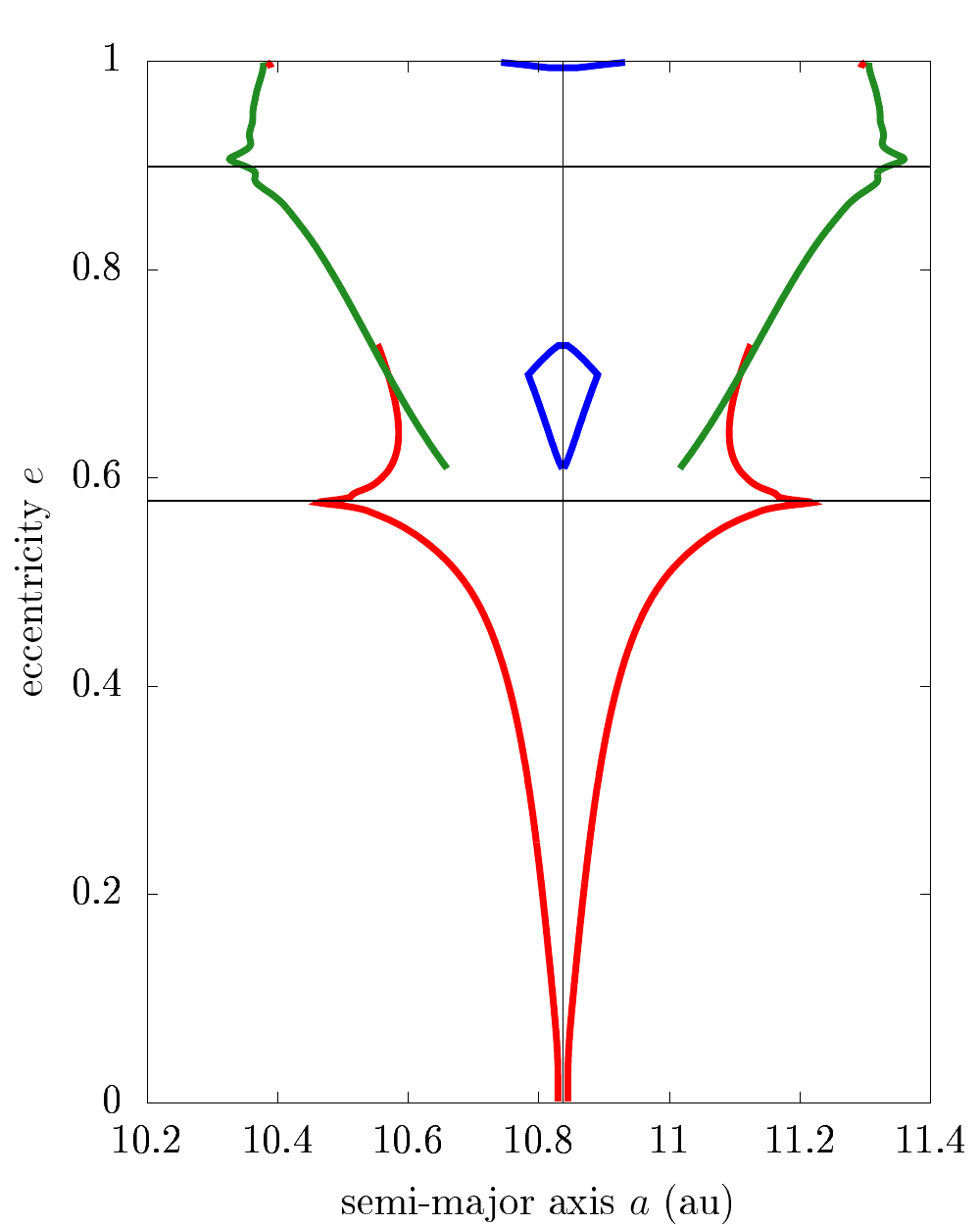}
      \caption{Same as Fig.~\ref{fig:Geminids} for the $1$:$3$ commensurability with Jupiter. The small body has $i=167^\circ$, $\omega=228^\circ$, and $\Omega=170^\circ$, similarly to the Leonids (see Table~\ref{tab:CILeo}).}
      \label{fig:Leonids}
   \end{figure}

   Figures~\ref{fig:Geminids}, \ref{fig:Draconids} and \ref{fig:Leonids} show the widths of the islands and sub-islands of resonance as a function of eccentricity for different commensurabilities $k_p$:$k$ with the Earth and Jupiter. These diagrams reveal various bifurcations where a sub-island shrinks and vanishes, leading to the disappearance of an inner separatrix. Similar diagrams have been described by \cite{Gallardo_2019} and \cite{Namouni-Morais_2020} in the Solar System, \cite{Gallardo-etal_2021} in the non-restricted two-planet case, and \cite{Malhotra-Zhang_2020,Malhotra-Chen_2023} in the planar case using a non-perturbative approach.

%% file: Results.tex
\section{Results}\label{sec:results}

All maps are drawn as a function of initial orbital elements, and final OFLI. We only present maps drawn in function of initial semi-major axis and initial eccentricity, or in function of initial semi-major axis and initial mean anomaly. These orbital elements are representative of the short-term dynamics of the particles (see e.g. Appendix~\ref{asec:widths}); hence, they directly reflect the level of short-term chaoticity that we are looking for in this article. We will use the names `mean anomaly map' and `eccentricity map' to describe the two maps for ease of reference, even though they are also drawn as a function of initial semi-major axis and final OFLI.

\subsection{Draconids}

\subsubsection{Mean-motion resonances and close encounters}

\begin{figure}
    \includegraphics[scale = 0.5]{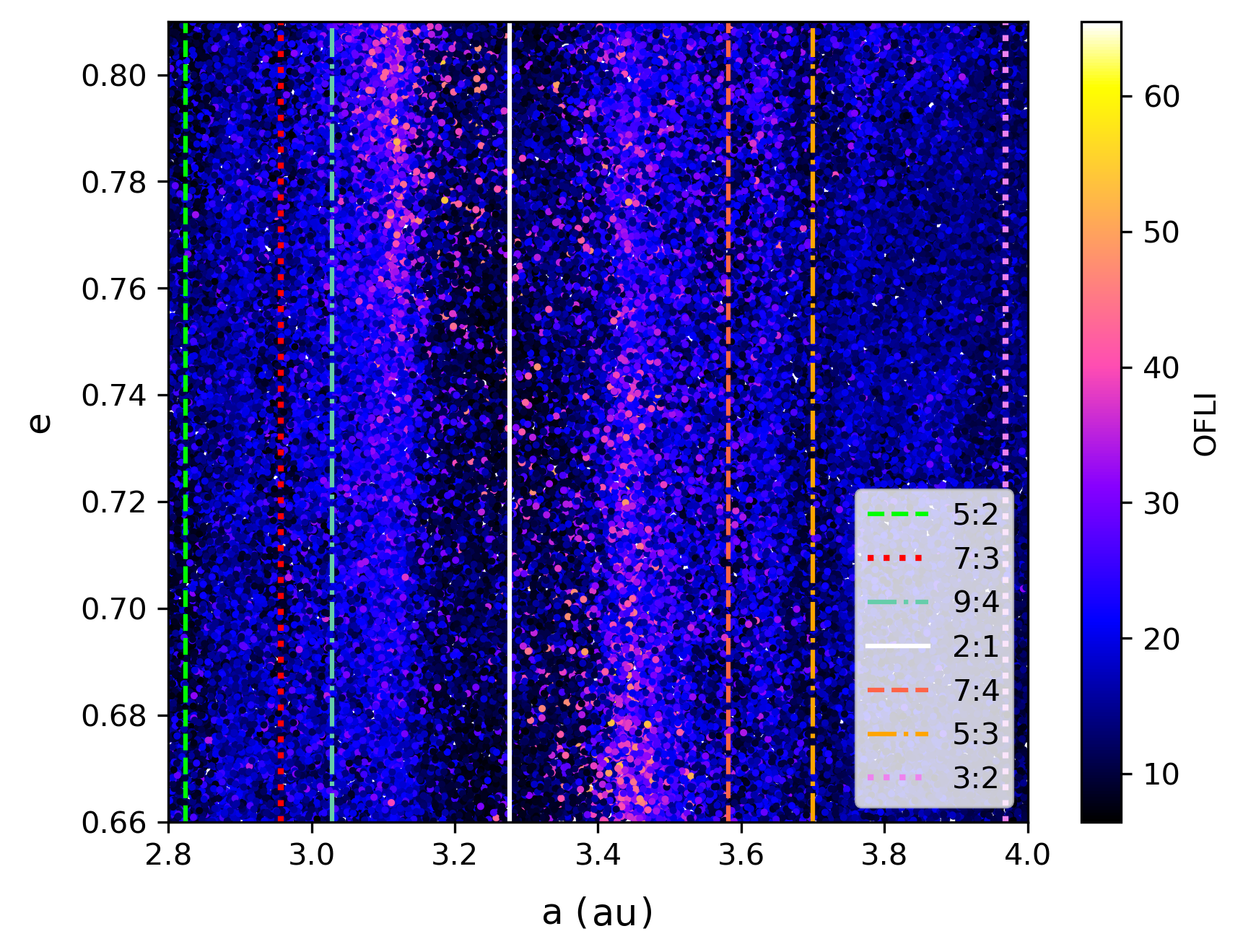}
    \includegraphics[scale = 0.5]{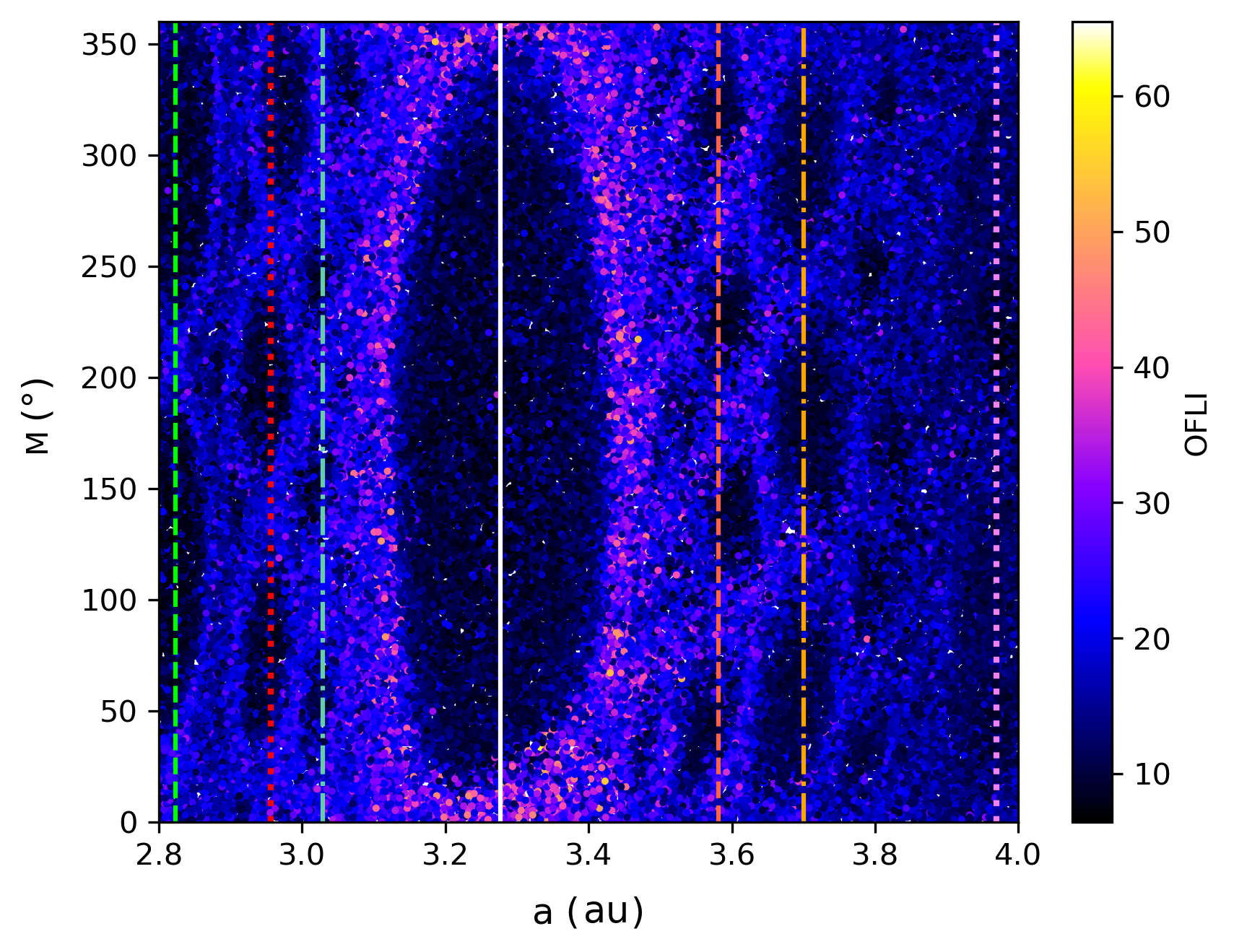}
    \caption{Eccentricity and mean anomaly maps drawn from the data set Draconids BIN10100, after the 1000 years of integration. The lines represent the resonances we identified (see text).}
    \label{fig:draco_reso}
\end{figure}

Maps from the BIN10100 set are presented in the Fig.~\ref{fig:draco_reso}. Here the NGFs are negligible, because they do not play a role for such large particles \citep[see also][]{Courtot_al_2023}.

We identify seven resonances with Jupiter in the mean anomaly map (see Table \ref{tab:draco_reso}). Only the largest (2:1) is undoubtedly visible in the eccentricity map. Those resonances are easily identified thanks to the number of lobes, which reveals the value of the denominator. For the largest resonance in particular, the separatrix around the resonance is characterised by highly chaotic trajectories (bright pink dots; high values of OFLI). It is indeed well known that chaos first appears at the separatrix in hamiltonian dynamics \citep{Poincare_1890}.

One may be tempted to relate the large width of the 2:1 resonance to its low `order' (which is traditionally defined as $2-1=1$). However, strictly speaking, this definition of the resonance order is only valid in the low-eccentricity, low-inclination paradigm. Here both the Draconids and Leonids have very large eccentricities and inclinations, so the traditional notion of resonance order loses its meaning and should be redefined (see e.g. \citealp{Morais_Namouni_2013,Namouni_Morais_2017,Morais_Giuppone_2012}). Therefore, we will avoid using this notion and instead will focus on the width of the MMRs, as visible in the maps or as computed semi-analytically.

\begin{table}
    \caption{Mean-motion resonances (MMRs) with Jupiter found in the Draconids maps.}
    \label{tab:draco_reso}
    \centering
    \begin{tabular}{c c}
    \hline \hline
         MMR & $a$ (au) \\
         \hline
         5:2 & 2.82364 \\ 
         7:3 & 2.95655 \\ 
         9:4 & 3.02911 \\ 
         2:1 & 3.27655 \\ 
         7:4 & 3.58161 \\ 
         5:3 & 3.70002 \\ 
         3:2 & 3.96926 \\ 
         \hline
    \end{tabular}
    \tablefoot{The central semi-major axis values of the resonances are computed using Eq.~\eqref{eq:a0}. No other MMRs with Jupiter were found visually in the Draconids maps.}
\end{table}


The least visible resonance is the 9:4 resonance, perhaps because of its closeness to the largest resonance, while itself being very thin. All resonances identified here were already detected in previous works, such as \cite{Fernandez_al_2014}.

Close encounters are detected when a particle is closer to the planet than its Hill radius. Even though we detected close encounters with Mercury (3 particles), Venus (1031), the Earth (1445), Mars (106), Saturn (19), Uranus (2) and Neptune (1), by far the most encounters happened with Jupiter (30397 particles). For this reason, as well as the interplay with the MMRs, our analysis below focuses only on close encounters with Jupiter..

\begin{figure}
    \includegraphics[scale = 0.5]{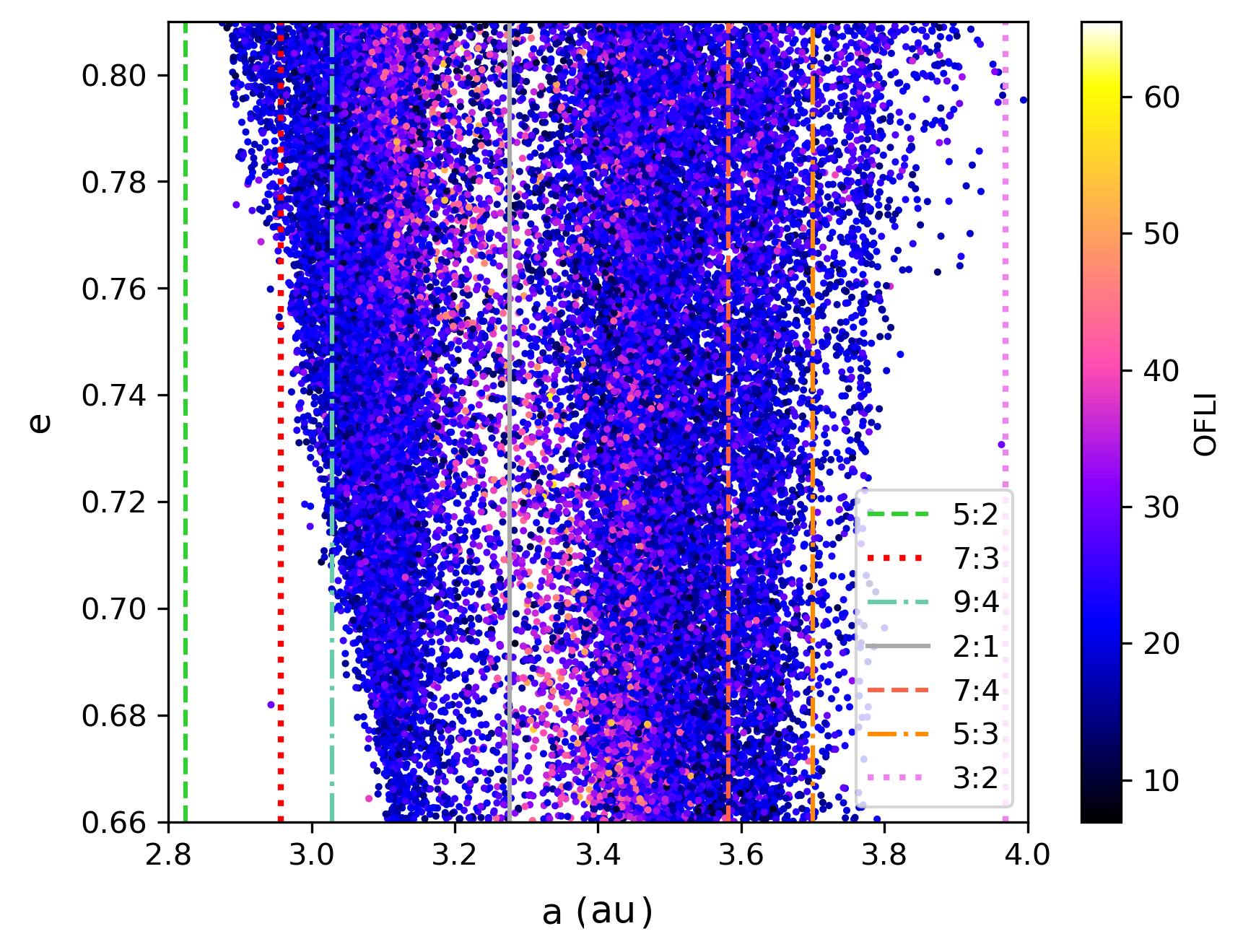}
    \includegraphics[scale = 0.5]{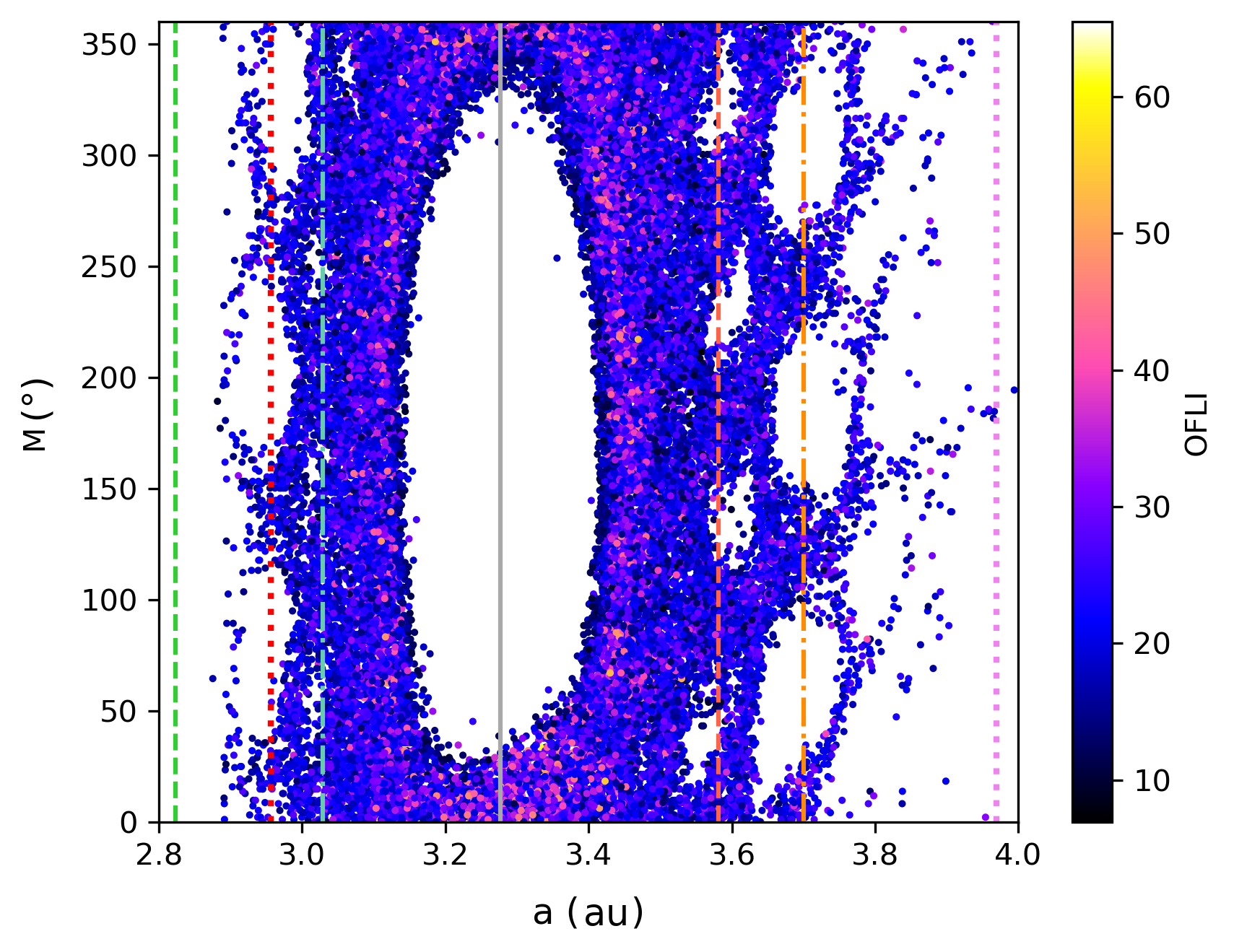}
    \caption{Same as Fig.~\ref{fig:draco_reso}, but only showing particles which encountered Jupiter during the integration, which means the white area shows the initial phase space that do not lead to a close encounters with Jupiter.}
    \label{fig:draco_renc}
\end{figure}

Fig.~\ref{fig:draco_renc} presents maps for BIN10100 once again, but this time only particles that encountered Jupiter during the integration are plotted.
The mean anomaly map shows clearly the effect of the MMRs combined with the close encounters: particles trapped inside the MMRs do not meet with Jupiter. Only the leftmost and rightmost MMRs (5:2 and 3:2) do not present such obvious features. There seems to be less close encounters in general for these semi-major axes, which means the effect of the capture is less visible.

\begin{figure}
    \includegraphics[scale = 0.5]{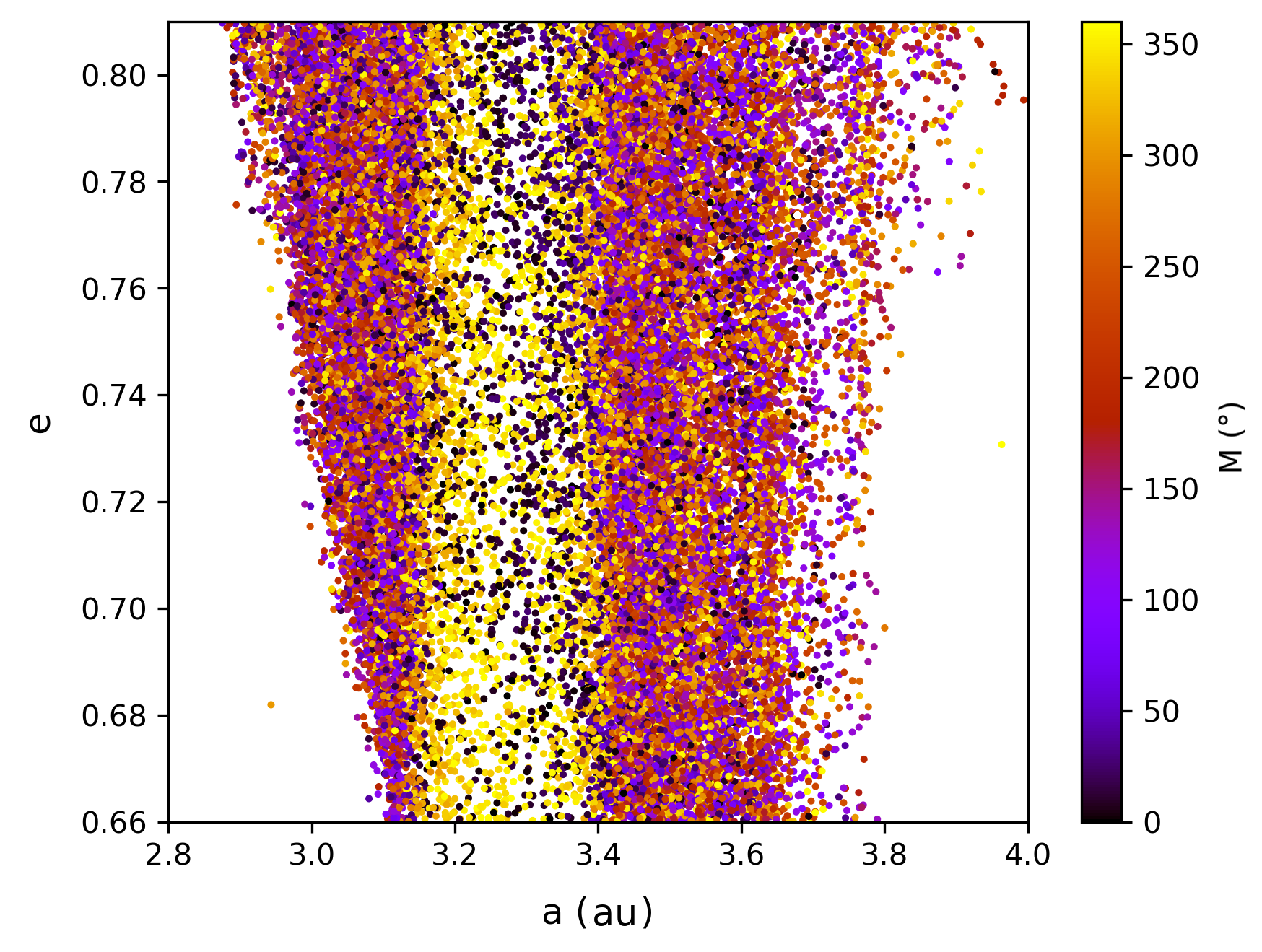}
    \caption{Same as Fig.~\ref{fig:draco_renc}, except that the colour scale shows the initial mean anomaly for each particle.}
    \label{fig:draco_aem}
\end{figure}

As for the eccentricity map, as before the only resonance undoubtedly visible is the 2:1. But there are encounters with Jupiter even inside this resonance. This is in fact the effect of the separatrix: the only encounters inside the resonance are linked with initial mean anomaly close to 0°[2$\pi$], where the separatrix lays. To verify this assumption, we drew another eccentricity map in Fig.~\ref{fig:draco_aem}, from particles that meet Jupiter, but the colour bar represents the value of the initial mean anomaly, instead of the OFLI. 

We were also able to verify that the position of the separatrix is linked with the initial position of Jupiter, in mean longitude. For initial conditions taken a few years earlier or later, the resonance islands on the mean anomaly maps appear shifted vertically.

\subsubsection{Effect of non-gravitational forces}

\begin{figure}
    \includegraphics[scale = 0.5]{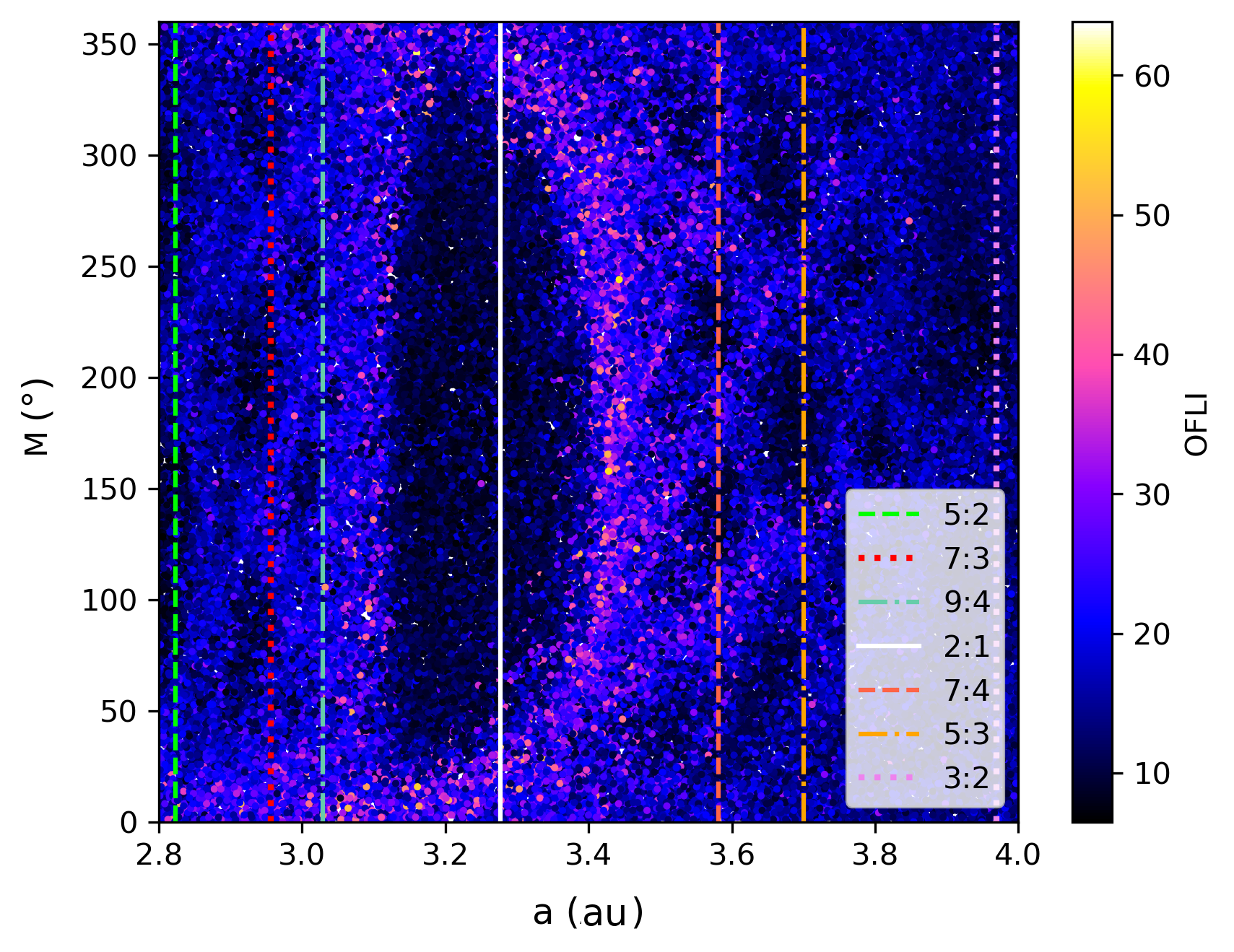}
    \caption{Mean anomaly map from the data set Draconids BIN00101.}
    \label{fig:draco_bin00101}
\end{figure}

In order to investigate the effect of NGFs, we studied the data sets with smaller radii. We drew maps from BIN110 and BIN011, but found no notable difference from the previous maps. We had to use the data set BIN00101 to visualise the effect of NGFs, and only the mean anomaly map revealed it (see  Fig.~\ref{fig:draco_bin00101}). The lobes are all fuzzier, especially for the 5:2, 7:3 and 9:4 MMRs. The main lobe (2:1 MMR) is also slightly distorted, losing its symmetry, and more chaos is visible on the right side of the separatrix than on its left side (compare with Fig.~\ref{fig:draco_reso}). The NGFs make the particles diffuse towards smaller semi-major axis, which means that particles initially on the left side of the separatrix can leave this chaotic zone thanks to the NGFs and reach zones where encounters with Jupiter will be less chaotic. On the other hand, particles on the right side of the separatrix are blocked from diffusing to the left because of the separatrix itself. Particles trapped on this right side of the separatrix will experience more chaos and more chaotic encounters with Jupiter.

These differences are much subtler than what we could have expected from our previous work on the Geminids, where the map from BIN011 revealed the destruction of the resonances involved because of the NGFs \citep{Courtot_al_2023}. However, two parameters might play a role in this difference: the type of orbit and the width of the resonances.

According to \cite{Liou_Zook_1997}, only the Poynting-Robertson drag could be responsible for the escape of the particles from the MMRs, if it happens. Geminids are characterised by a smaller perihelion distance, a smaller semi-major axis and thus a greater velocity than Draconids. Therefore we can expect a highly different effect of the Poynting-Robertson drag: it will probably be too weak to make the small Draconids diffuse out of the MMR.

\cite{Liou_Zook_1997} showed that the drift rate in semi-major axis due to this force can be written:
\begin{equation}\label{eq:drift}
    \left( \frac{da}{dt} \right)_{PR} = \frac{-1.35 \beta \mu_\sun}{c} \frac{2 + 3 e^2}{a (1 - e^2)^{3/2}},
\end{equation}
with $\beta$ the ratio of the radiation pressure force and the Sun gravitational force, $\mu_\sun$ the gravitational constant of the Sun and $c$ the speed of light. 

We also have \citep{Burns1979}:
\begin{equation}
    \beta = \frac{3}{4} \frac{S_0 R_0^2}{\mu_\sun c} \frac{1}{\rho r},
\end{equation}
with $\rho$ the density of the particle, $r$ its radius and $S_0$ the solar flux at $R_0 = 1$~au, its value is $1.37 \text{kW}\ \text{m}^{-2}$ \citep{Cox_2000}.

We choose a radius of $0.3$~mm, which corresponds to the largest radius at which a Geminid particle can escape the MMR through diffusion from Poynting-Robertson \citep{Courtot_al_2023}.


We apply this equation to the Geminids ($a = 1.275$~au, $e = 0.875$) and we obtain $-1.53 \times 10^{-12} \text{au}\ \text{s}^{-1}$ and for the Draconids ($a = 3.4$~au, $e = 0.735$), $-1.75 \times 10^{-13} \text{au}\ \text{s}^{-1}$. This would mean that over 1000 years, the drift in semi-major axis due to Poynting-Robertson would be of $-0.048$~au for the Geminids and of $-0.006$~au for the Draconids.

\begin{figure}
    \includegraphics[scale = 0.5]{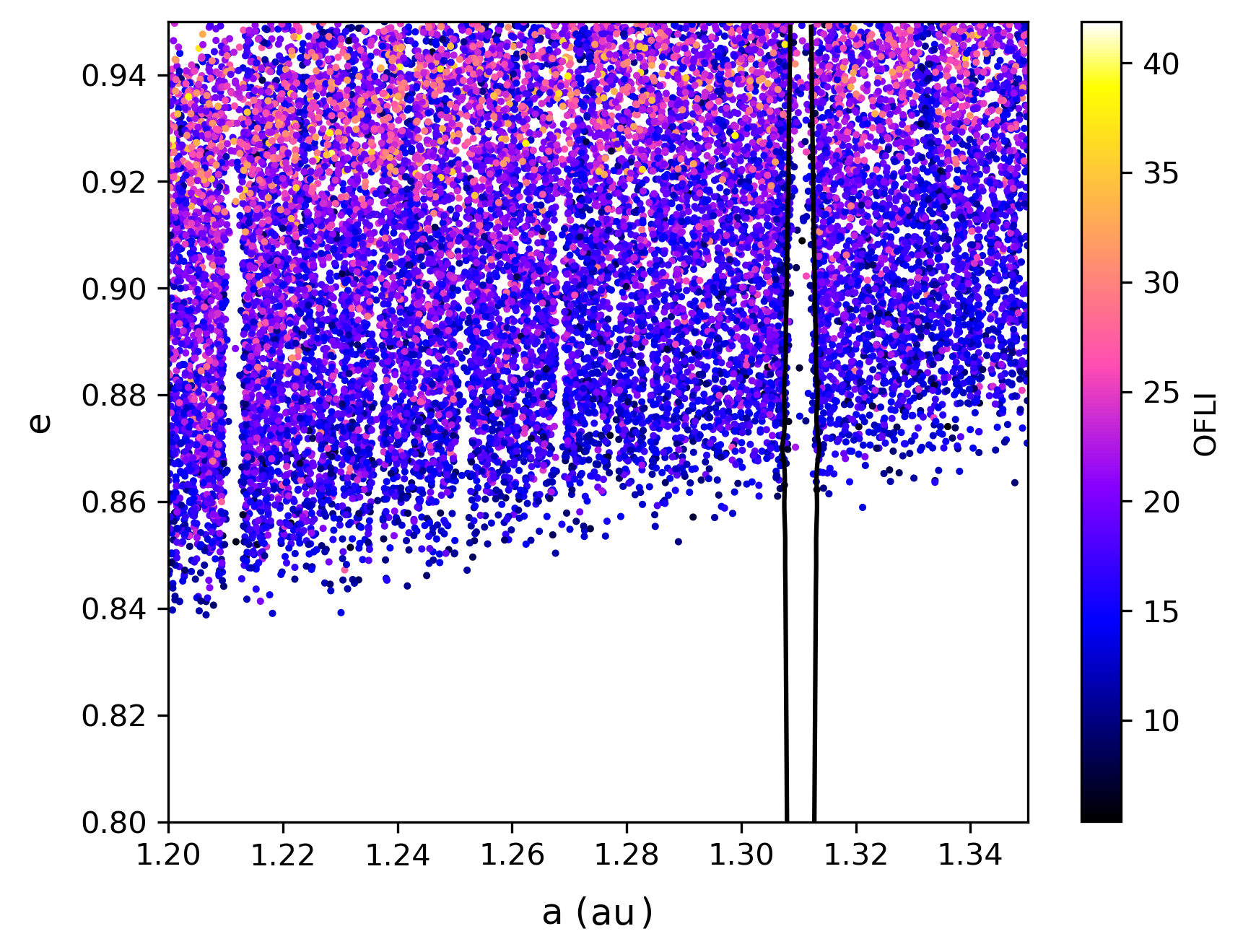}
    \includegraphics[scale = 0.5]{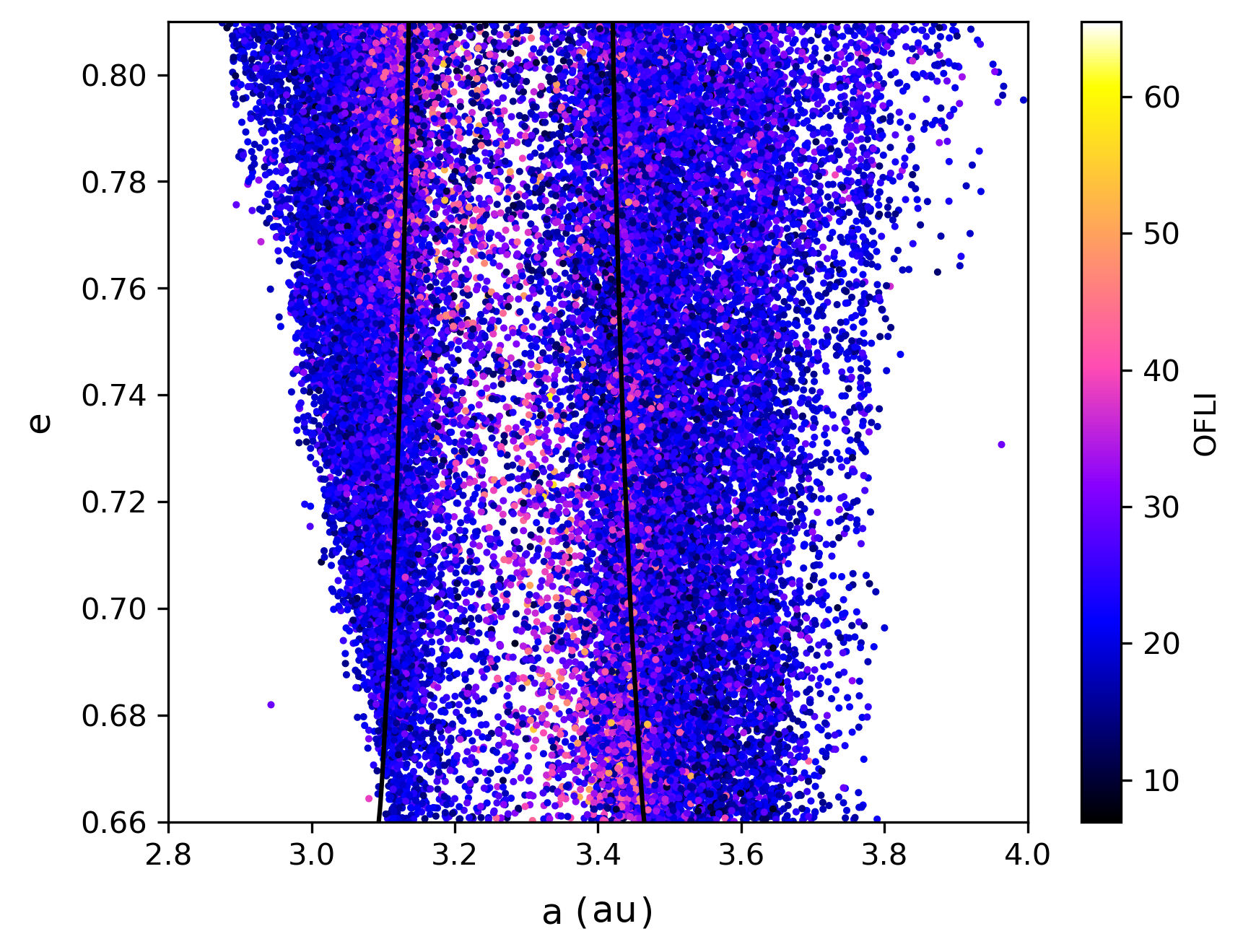}
    \caption{Width of the MMR for the Geminids (2:3 with the Earth; top) and for the Draconids (2:1 with Jupiter; bottom).  The first map is adapted from \cite{Courtot_al_2023}: we only added the newly computed width of the MMR on a map presented therein (radius between 10 and 100~mm, integration during 1000 years, all particles plotted encountered the Earth). In the second case, we added the eccentricity map from BIN10100, with only particles that meet with Jupiter: the resonance is between the two dark lines.}
    \label{fig:comp_draco_gem}
\end{figure}

Furthermore, the widest MMR for the Geminids is 2:3 with the Earth, which is much thinner than the widest MMR for the Draconids. We computed the width of both of those MMRs using the method described in Sect.~\ref{sec:methwidth}.
We used typical values of the orbital elements of both meteoroid streams as parameters (see Figs.~\ref{fig:Geminids} and \ref{fig:Draconids}). Figure \ref{fig:comp_draco_gem} shows the comparison we obtained. For the range in eccentricity considered, the maximum width for the Geminds main MMR is $4.76 \times 10^{-4}$~au, compared to $5.12 \times 10^{-1}$~au for the Draconids main MMR.

In summary, not only the Poynting-Robertson drag is more efficient for the Geminids, but more importantly, the width of the MMR that can trap Geminid particles is much smaller than the width of the MMR that can trap Draconids. This explains why, even at a low radius, Draconids tend to stay captured in the MMR, contrary to what we observed with the Geminids.




\subsection{Leonids}

\subsubsection{Mean-motion resonances and close encounters}

\begin{figure}
    \includegraphics[scale = 0.5]{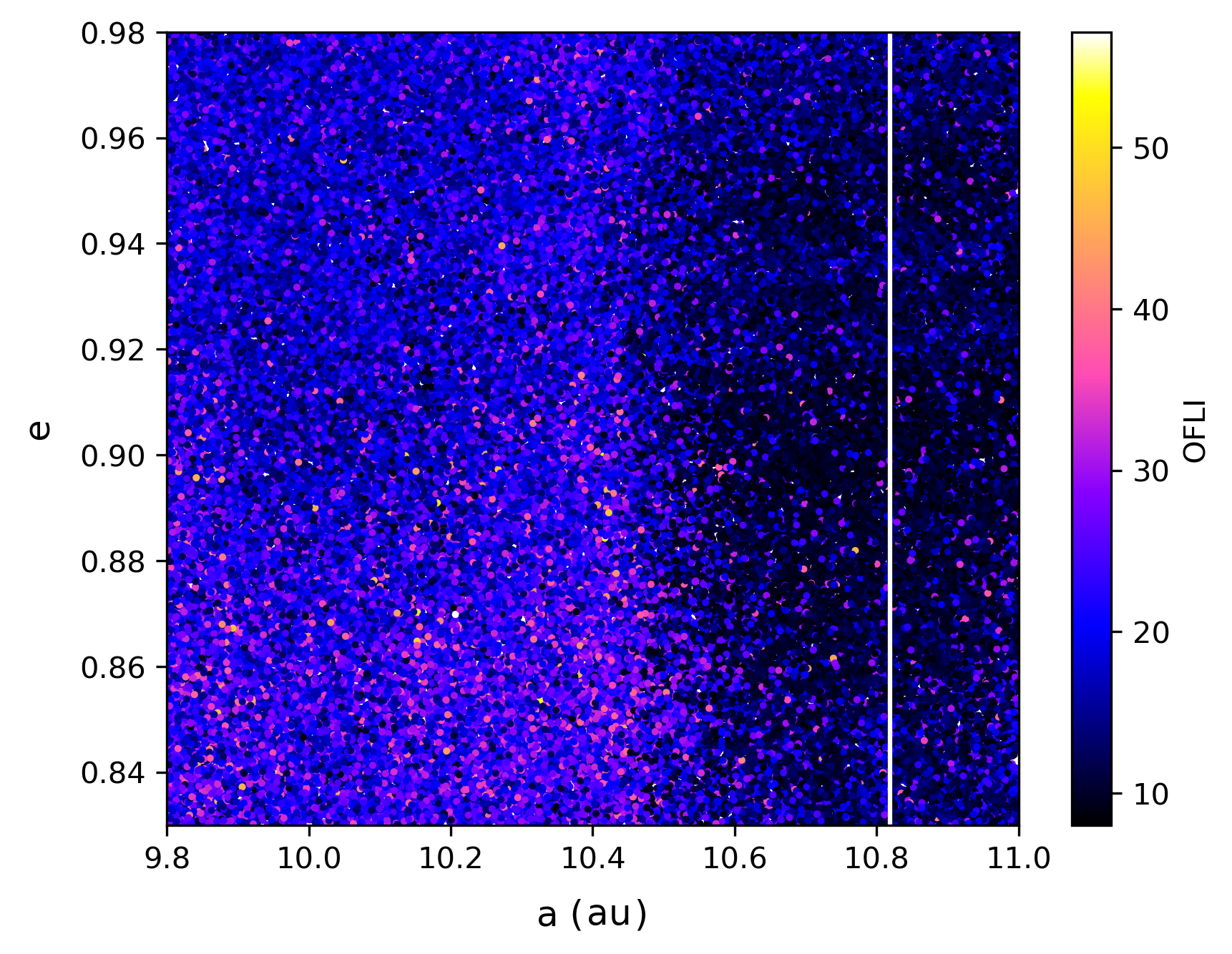}
    \includegraphics[scale = 0.5]{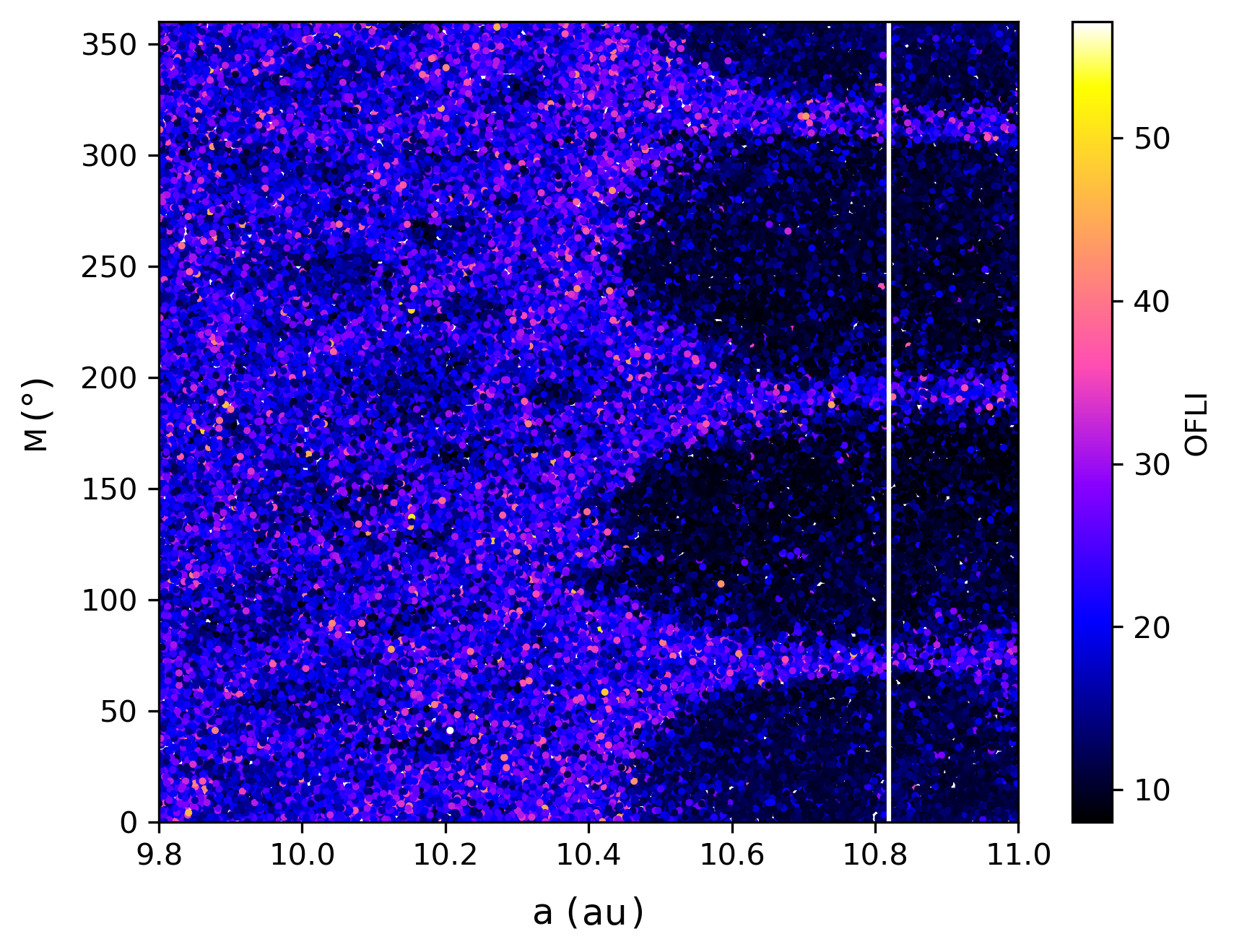}
    \caption{Eccentricity and mean anomaly maps from the data set Leonids BIN10100, after the 2000 years of integration. The white line represents the position of the 1:3 MMR with Jupiter.}
    \label{fig:leo_reso}
\end{figure}

As with the Draconids, we first drew maps from the BIN10100 data set. In the Fig.~\ref{fig:leo_reso}, a MMR with Jupiter is detected: the 1:3 at 10.83681~au. This MMR is visible in both eccentricity and mean anomaly maps, with the three lobes present in the latter.

\begin{figure}
    \includegraphics[scale = 0.5]{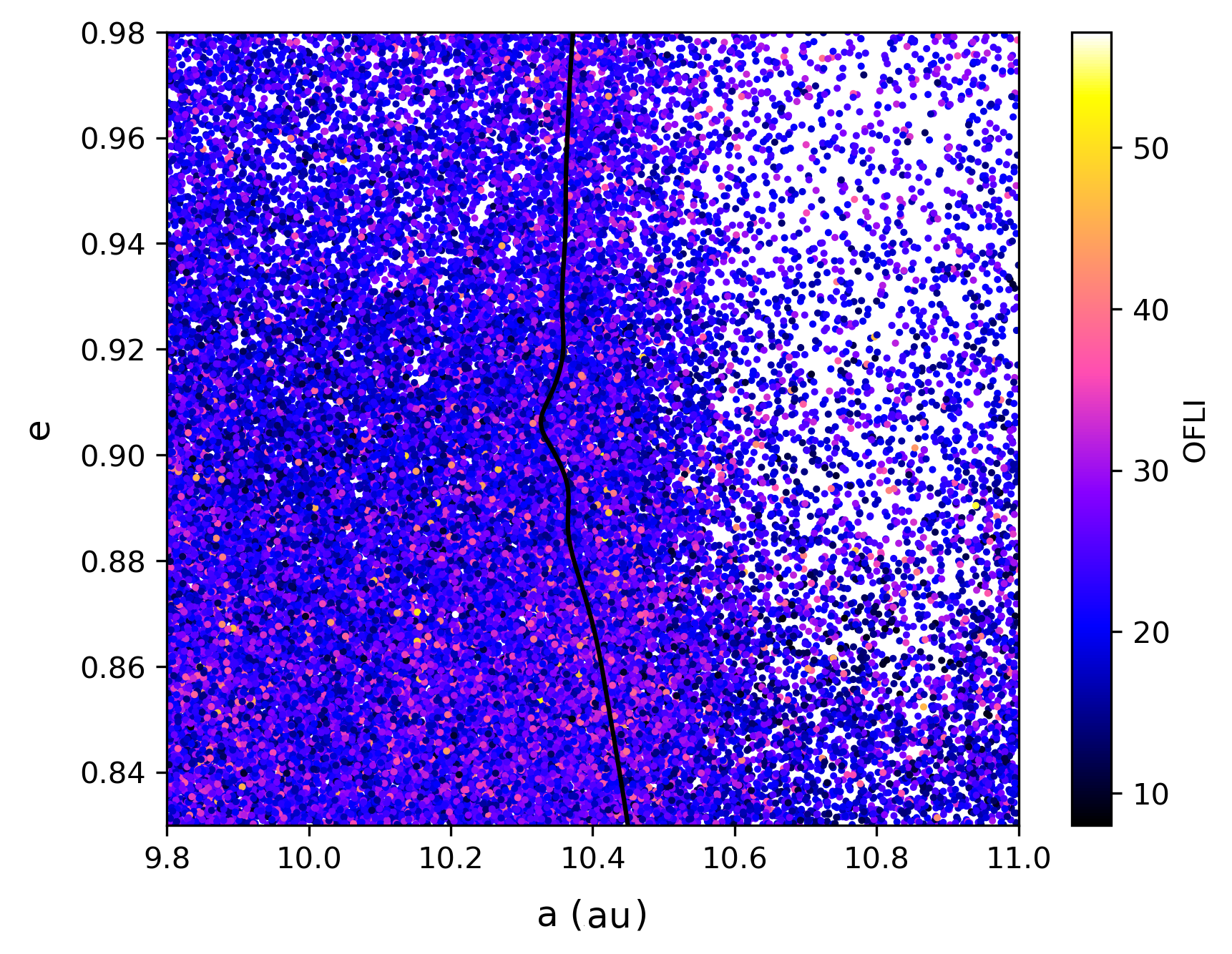}
    \includegraphics[scale = 0.5]{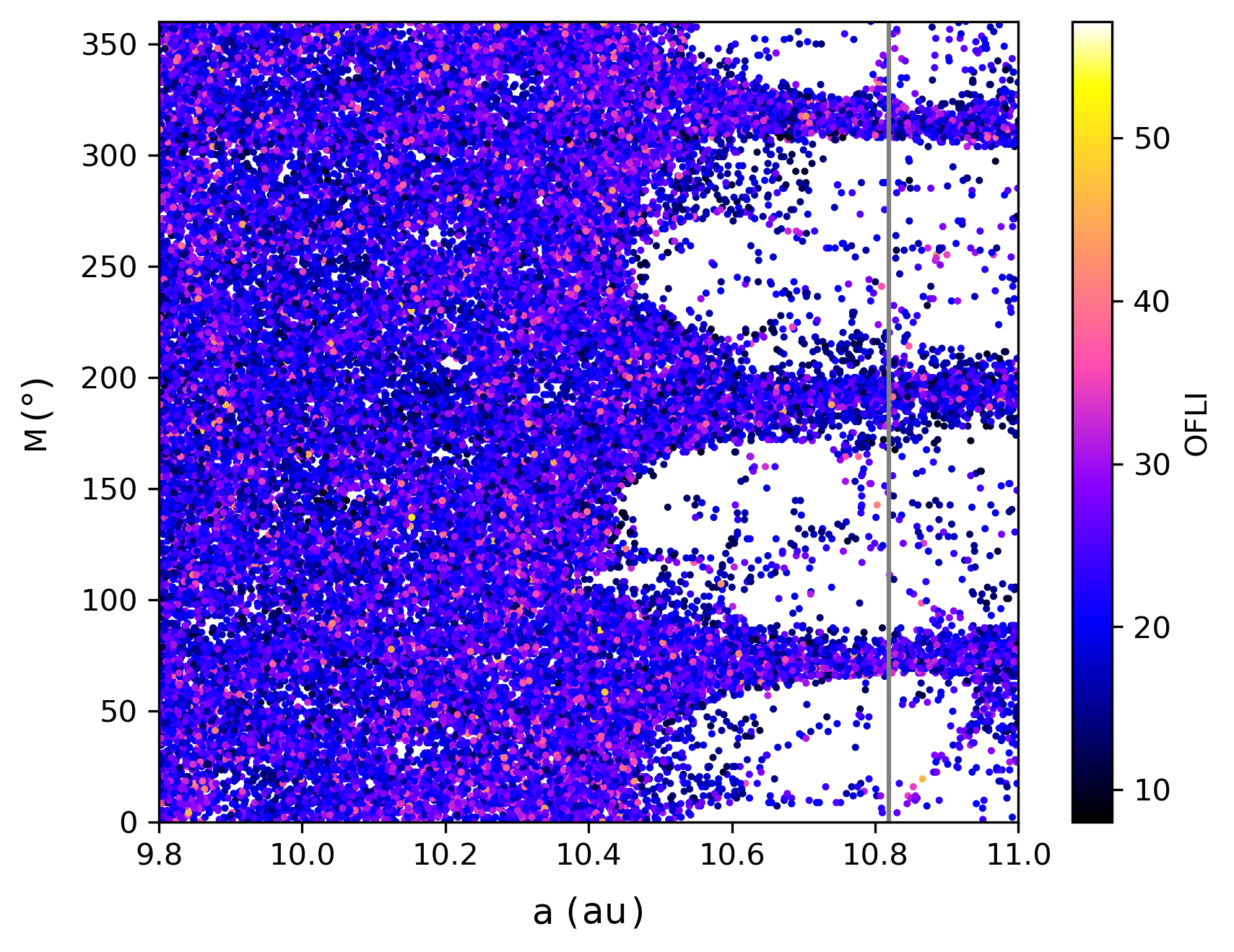}
    \caption{Eccentricity and mean anomaly maps from the data set BIN10100. We only plotted particles that met with Jupiter at least once during the 2000 years of integration. On the eccentricity map, we plotted the width of the MMR (1:3 with Jupiter). The resonance is too wide for our set of initial conditions: the black line represents the limit of the MMR on its left side (see Fig.~\ref{fig:Leonids} for the full resonance width).}
    \label{fig:leo_rencjup}
\end{figure}

A similar mechanism to the one presented for the Draconids takes place, where the majority of close encounters with Jupiter occur for particles either outside the MMR or on the separatrix. This is proven in Fig.~\ref{fig:leo_rencjup} by drawing maps from particles that underwent at least one close encounter with Jupiter during the integration. The eccentricity map shows there are less encounters with Jupiter at high eccentricity, even outside the MMR, and, inside the MMR, more encounters for lower eccentricities.

\begin{figure}
    \includegraphics[scale = 0.5]{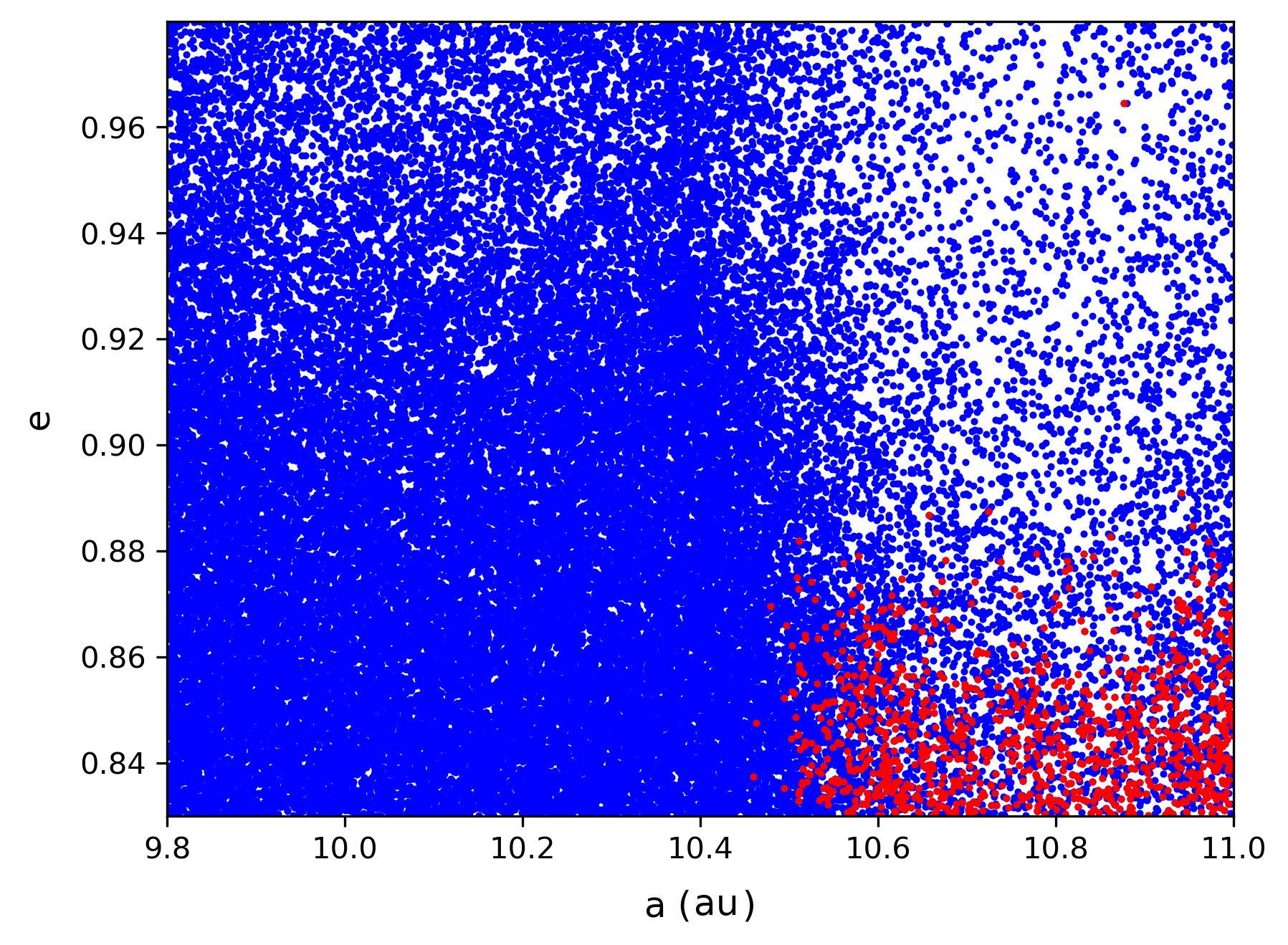}
    \includegraphics[scale = 0.5]{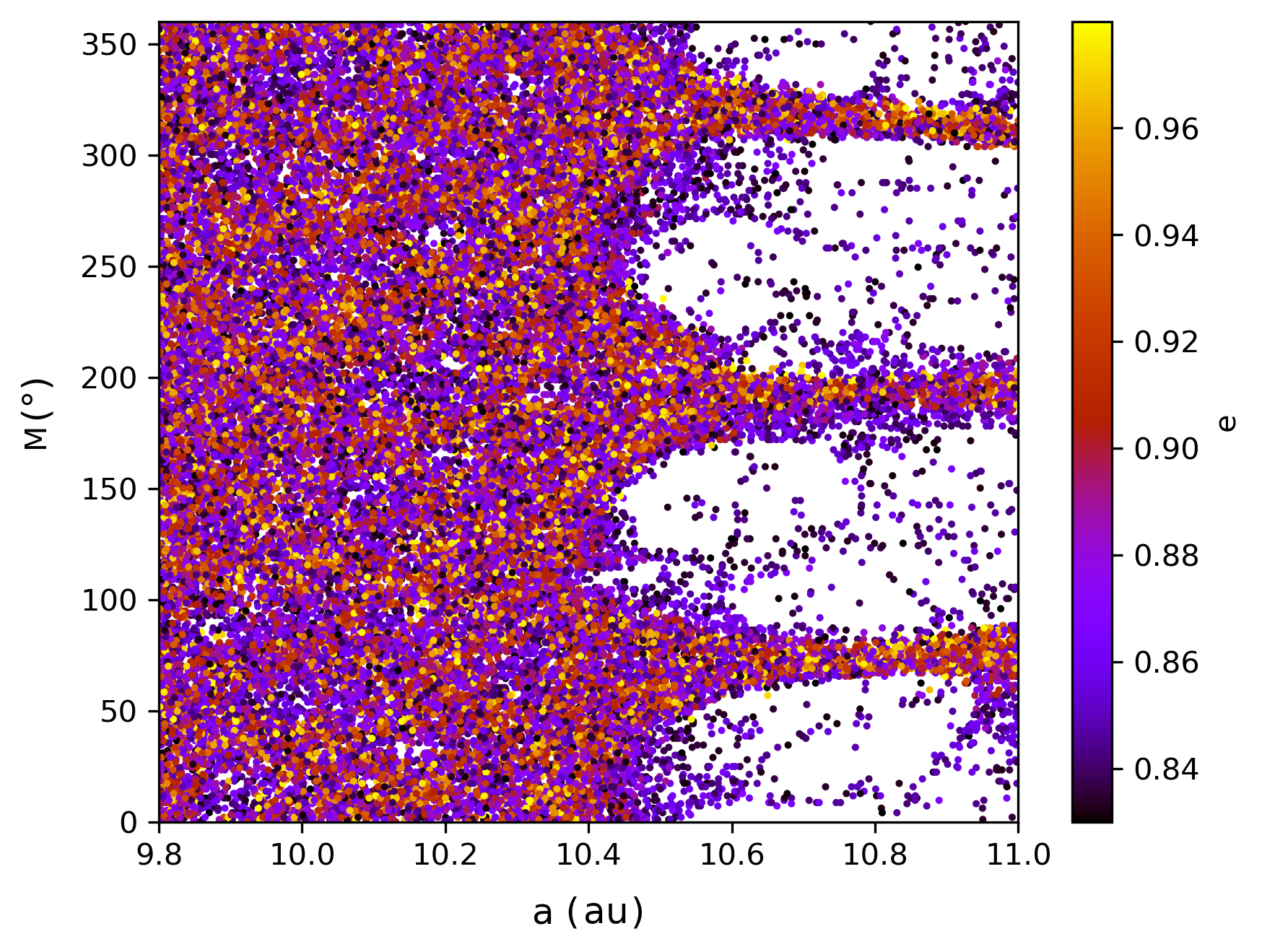}
    \caption{Map in mean anomaly and in eccentricity from the BIN10100 data set. We only plotted particles that encountered Jupiter. The eccentricity map presents 2 colours: red for particles from inside the lobes and blue for the rest. The colour bar from the mean anomaly map gives the value of the eccentricity.}
    \label{fig:leo_aem}
\end{figure}

Contrary to the Draconids, there are some close encounters with Jupiter for particles inside the lobes. Those particles either come from the separatrix region or have a relatively lower initial eccentricity ($\lesssim 0.87$), as it can be seen in Fig.~\ref{fig:leo_aem}. 

Amongst particles that do encounter Jupiter from inside the lobes, with a relatively small initial eccentricity, most met with Saturn during the integration, before meeting Jupiter. It seems those close encounters shifted their orbits enough to get them out of the lobes and therefore able to meet with Jupiter. 

\begin{figure}
    \includegraphics[scale = 0.5]{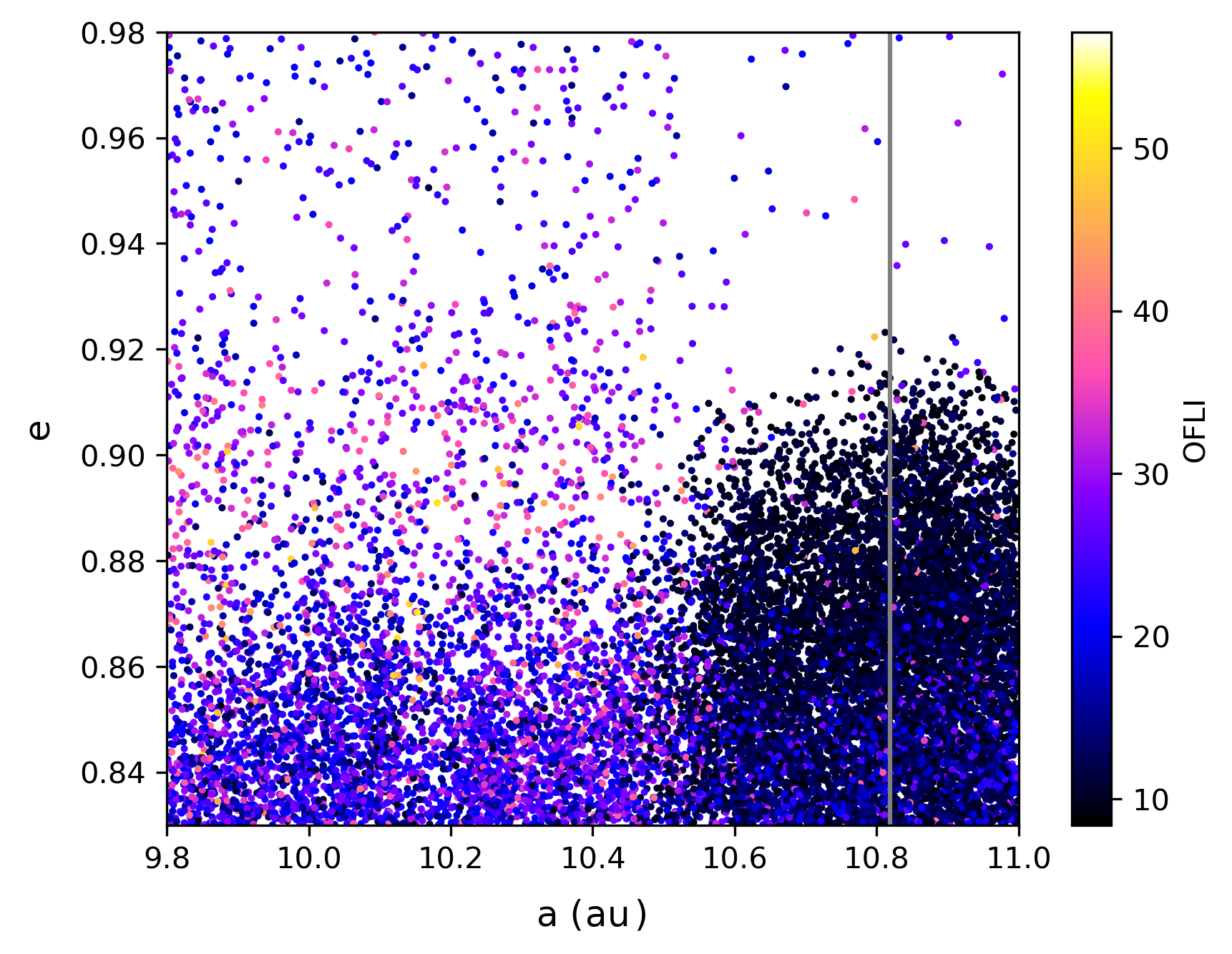}
    \includegraphics[scale = 0.5]{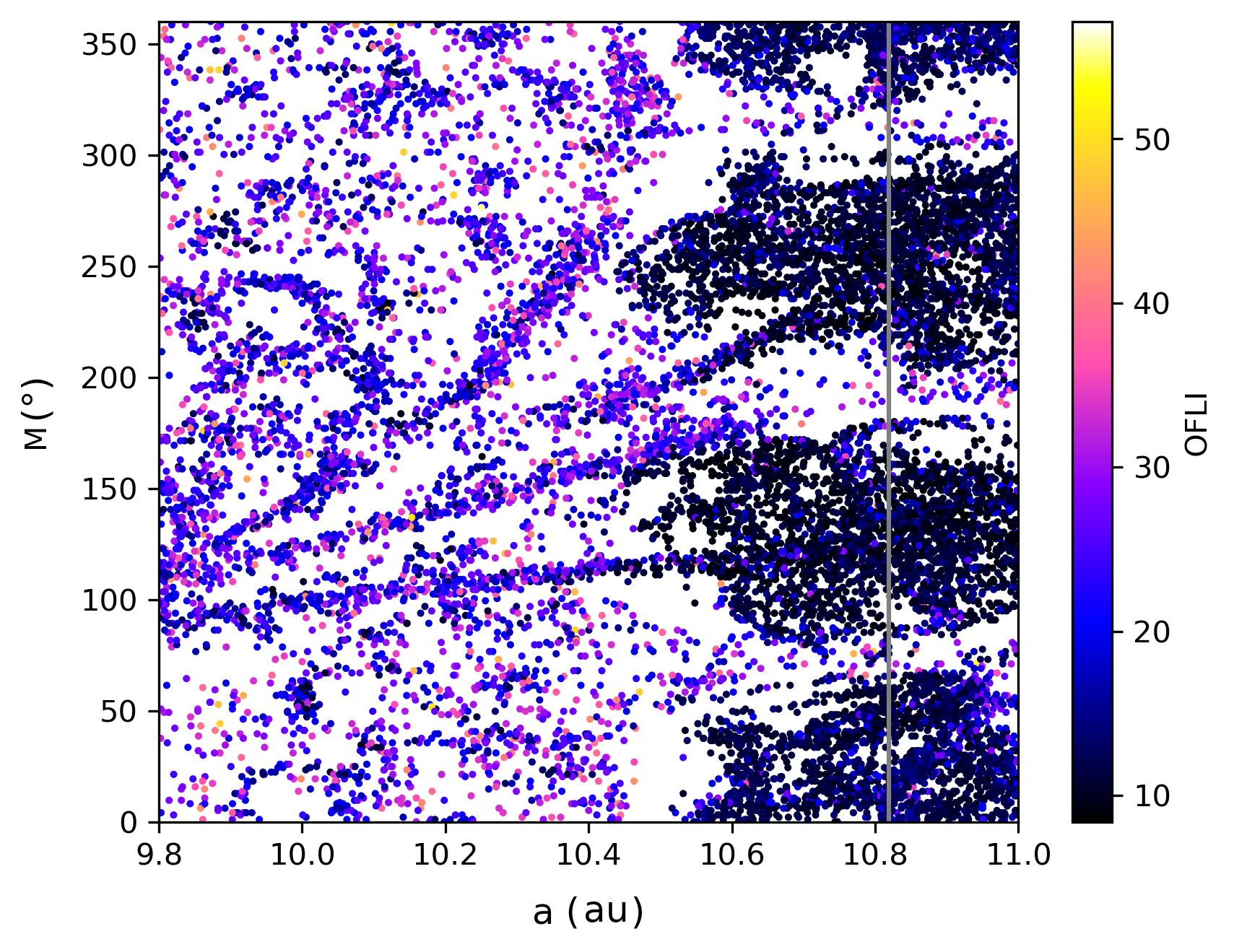}
    \includegraphics[scale = 0.44]{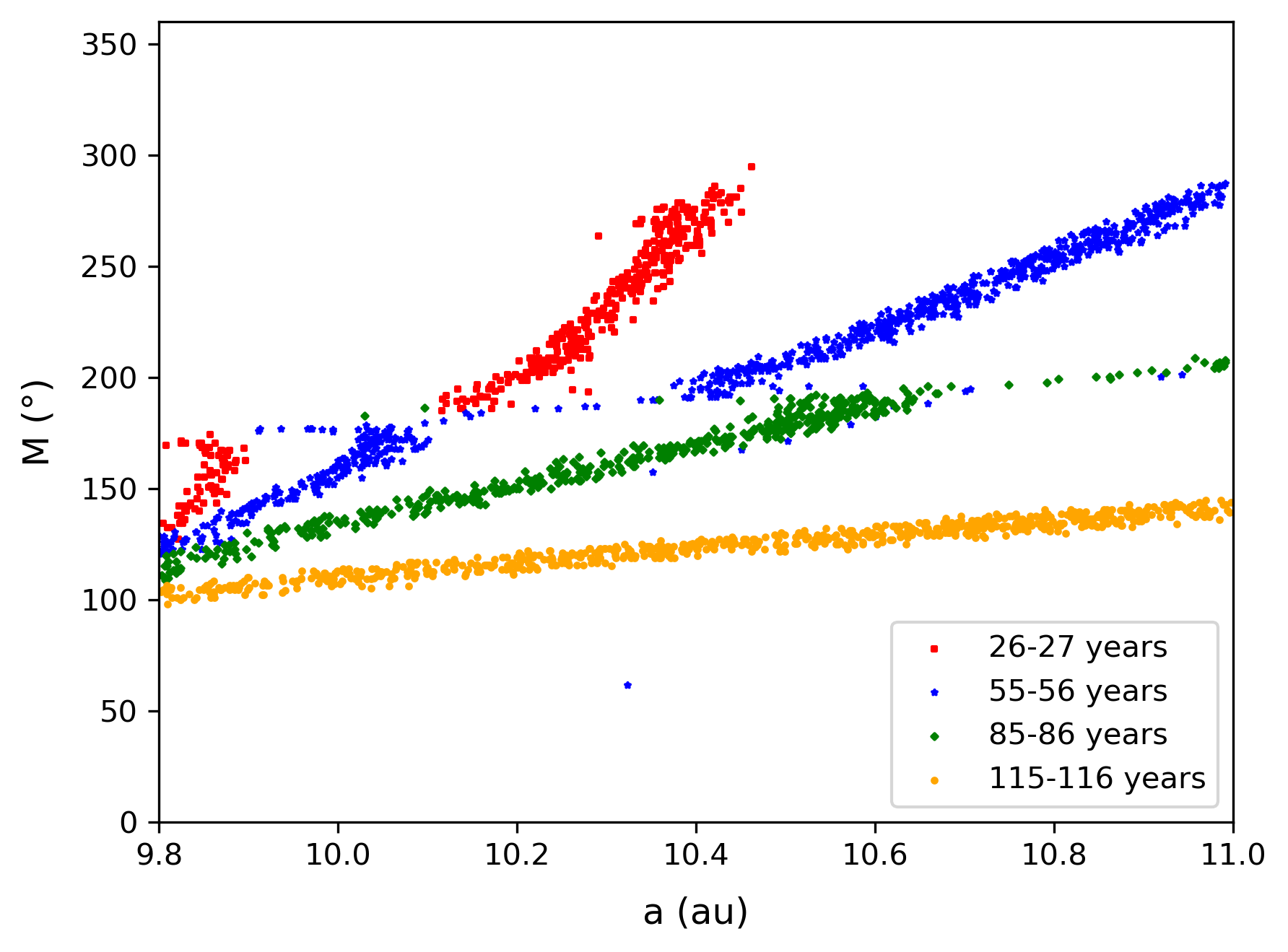}
    \caption{Eccentricity and mean anomaly maps from the BIN10100 data set. We only plotted particles that encountered Saturn. On the right-hand side of the mean anomaly map, the lobes are visible, while in the left-hand side, several lines appear. The last figure shows at which time those lines appear. Encounters outside the lines happen after 116 years of integration.}
    \label{fig:leo_rencsat}
\end{figure}

Close encounters with Saturn are shown in Fig.~\ref{fig:leo_rencsat}, where we drew our maps from particles that had a close encounter with Saturn. As expected, the eccentricity map reveals that most close encounters with Saturn happen for initially relatively low eccentricity, as these particles only need minimal orbital variations to reach an orbit-crossing  configuration.

The mean anomaly map, on the other hand, shows some interesting characteristics: lines unlike any other features seen so far. We were able to confirm that each line is created as time passes, when the orbits shift enough to create a new line (see legend of the figure). In other words, these lines represent the evolving relation between semi-major axis and mean anomaly necessary to encounter Saturn.

\begin{figure}
    \includegraphics[scale = 0.5]{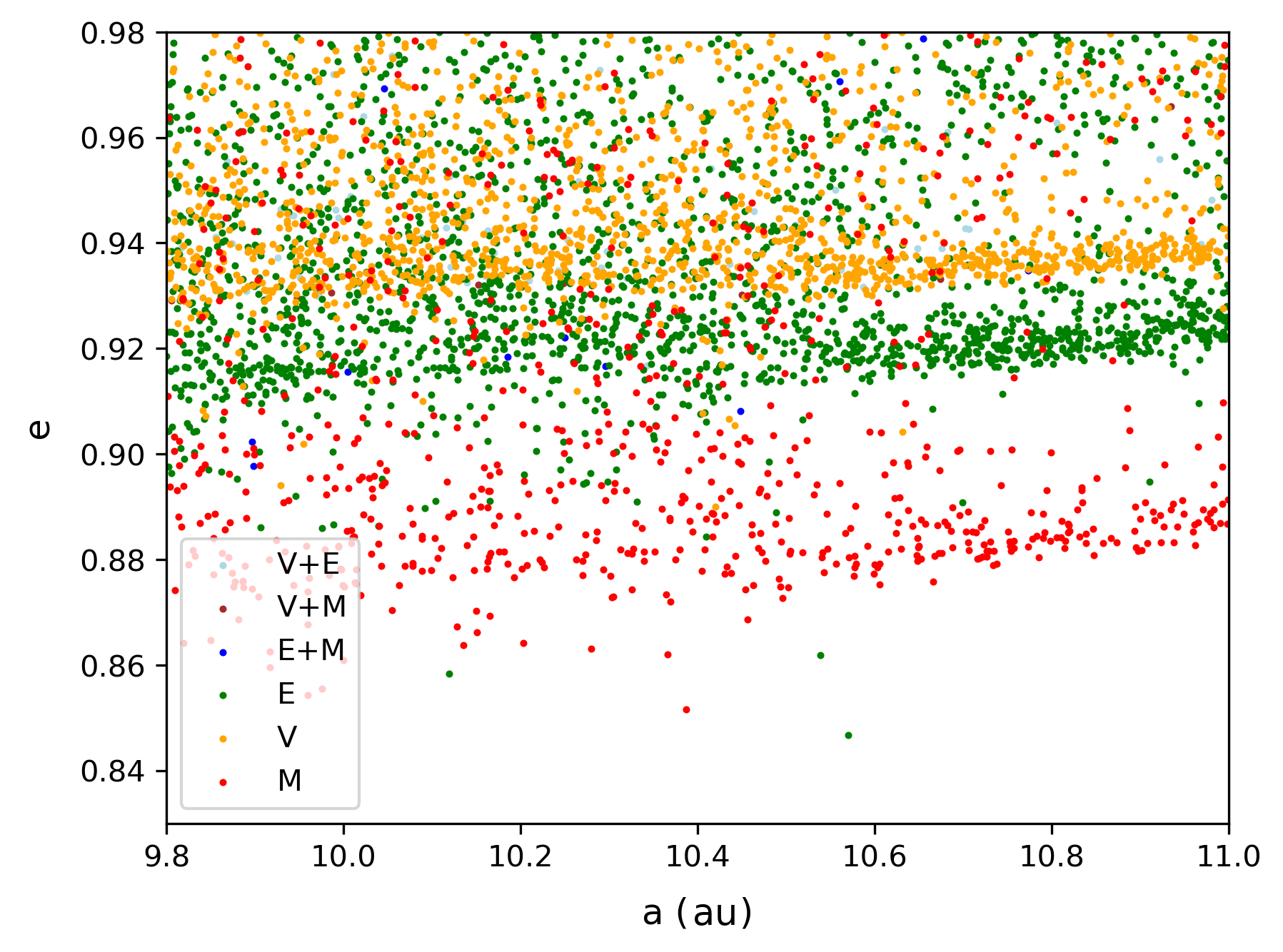}
    \caption{Map in eccentricity from the data set BIN10100. We plotted particles that met with Venus (orange), the Earth (green) or Mars (red). We also plotted particles that meet both Venus and the Earth (light blue), both Venus and Mars (brown), as well as both the Earth and Mars (dark blue), although those last three cases are not as common as the first three ones. The OFLI (not visible here) reaches similar values to the general map, its maximum being only slightly smaller.}
    \label{fig:leo_ssi}
\end{figure}

We also studied close encounters with the planets from the inner solar system in Fig.~\ref{fig:leo_ssi}. Those encounters happen mostly above a limit in mean anomaly, which depends on the planet considered, and is linear as a function of semi-major axis. These lines are very visible inside the MMR but get disturbed outside of it, most probably by close encounters with Jupiter. 

\subsubsection{Effect of non-gravitational forces}
\begin{figure}
    \includegraphics[scale = 0.5]{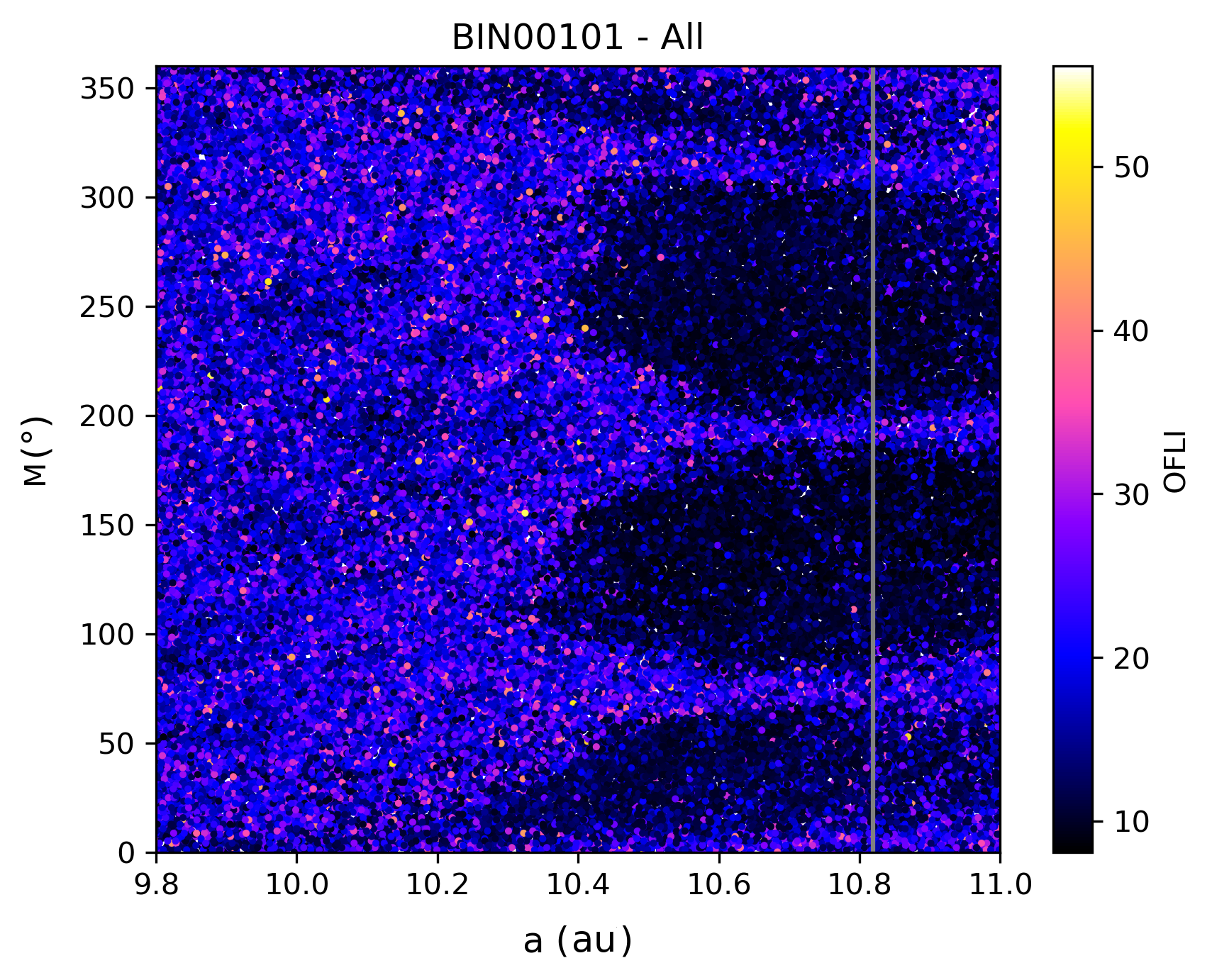}
    \includegraphics[scale = 0.5]{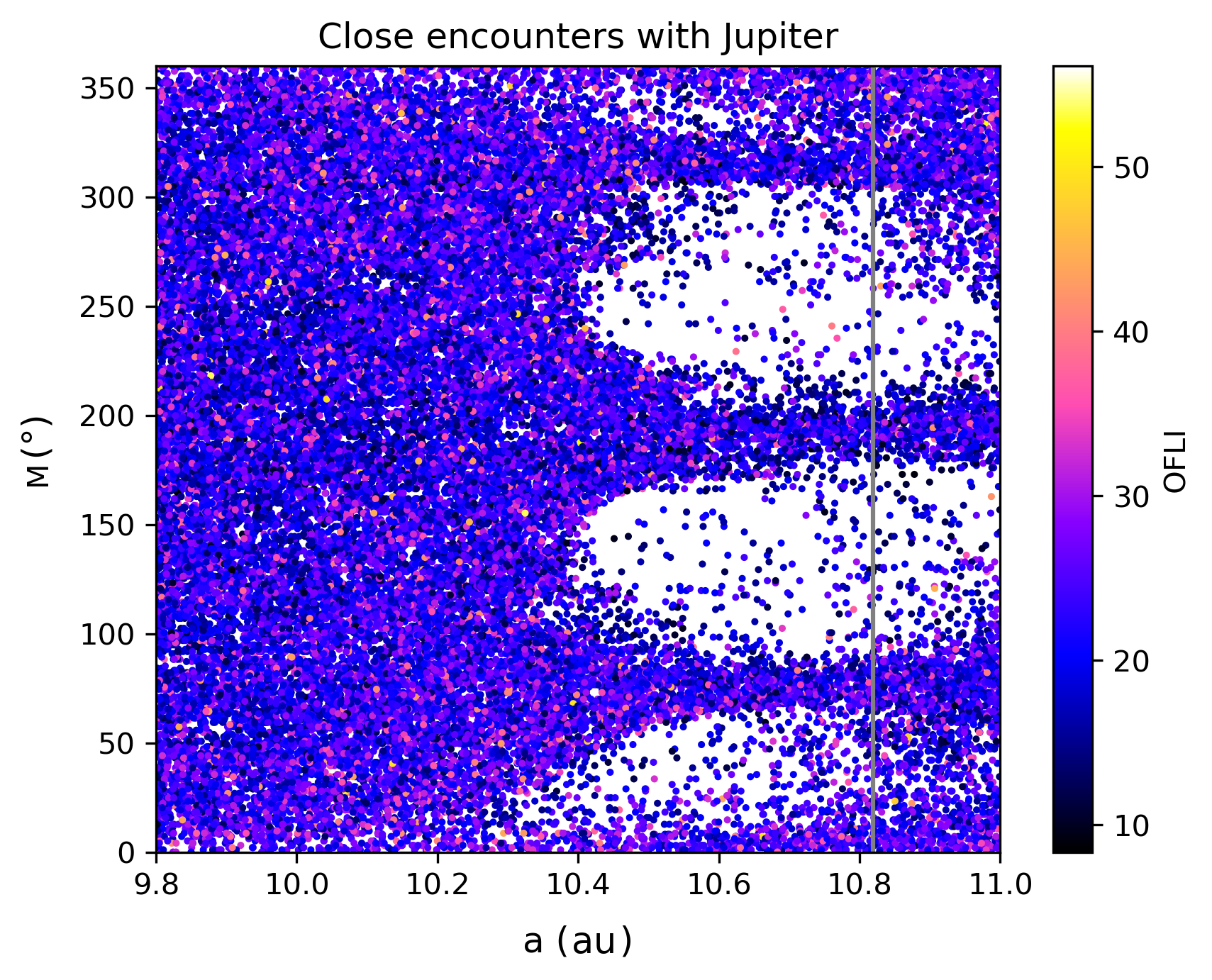}
    \caption{Maps in semi-major axis and mean anomaly from the data set BIN00101. One map is using all the particles, while the other one only shows particle that undergo a close encounter with Jupiter.}
    \label{fig:leo_fng}
\end{figure}

As with the Draconids, we had to use the data set with the smallest particles (BIN00101), in order to investigate the impact of NGFs. Even then, the only difference visible in Fig.~\ref{fig:leo_fng} is a bar of particles with high chaoticity, at an initial mean anomaly around $0^{\circ}$. In the same figure, we also drew a map from BIN00101 with only the particles that met with Jupiter during the integration: the bar clearly comes from close encounters with Jupiter. A combination of the initial position of Jupiter and the effect of NGFs, particularly strong at perihelion, are probably the origin of this new bar.

Once again, we expect the weaker Poynting-Robertson drag to be responsible for the difference in the Geminids and Leonids dynamics at such small radii. We apply the same equation \ref{eq:drift} to the Leonids (using $a = 10.4$~au and $e = 0.905$) and we obtain a velocity of drift in semi-major axis of $-2.86 \times 10^{-13}\text{au}\ \text{s}^{-1}$ , which would translate to a drift of $-0.009$~au over 1000 years. Those values are very similar to those previously obtained for the Draconids. 

We also compared the width of the 1:3 MMR with Jupiter playing a role for the Leonids with the 2:3 MMR with the Earth of the Geminids. Using representative values of the orbital elements (see Fig.~\ref{fig:Leonids}), we obtained the eccentricity map of Fig.~\ref{fig:leo_rencjup}, where the computed width of the MMR fits well with the detected MMR (for comparison of widths, see Fig.~\ref{fig:comp_draco_gem}.). 
The maximum width for the Leonids is of $2.61 \times 10^{-1}$~au, in the considered range of eccentricity, which is much greater than $4.76 \times 10^{-4}$~au for the Geminids. Once again, the combination of a weaker NGF and a larger MMR means that the dynamics of small particles is very similar to the behaviour of large particles.

%% file: Annexe.tex
\section{Computation of the semi-averaged resonant Hamiltonian function}\label{asec:widths}

   In Section~\ref{sec:methwidth}, we compute the widths of mean-motion resonances by a semi-analytical approach valid for arbitrary eccentricities and inclinations. Here, we detail the methodology used to average the Hamiltonian function and express it in a form that naturally leads to the adiabatic approximation.
   
   \subsection{Hamiltonian function in osculating coordinates}
   Using the notations of Sect.~\ref{sec:methwidth}, the Hamiltonian function describing the instantaneous dynamics of a small body perturbed by $N$ planets can be written as $\mathcal{H} = \mathcal{H}_0 + \varepsilon\mathcal{H}_1$ with
   \begin{equation}\label{eq:Hfirst}
      \left\{
      \begin{aligned}
         \mathcal{H}_0 &= -\frac{\mu}{2a} + \sum_{j=1}^{N}n_j\Lambda_j\,, \\
         \varepsilon\mathcal{H}_1 &= - \varepsilon\sum_{j=1}^{N}\mu_j\left(\frac{1}{\|\mathbf{r}-\mathbf{r}_j\|}-\mathbf{r}\cdot\frac{\mathbf{r}_j}{\|\mathbf{r}_j\|^3}\right)\,.
      \end{aligned}
      \right.
   \end{equation}
   The variable $\Lambda_j$ is the momentum conjugate to the mean longitude $\lambda_j$ of planet~$j$. Using these notations, $\mu$ and $\mu_j$ have the same order of magnitude, while $\varepsilon\ll 1$ is used to materialise the small size of the perturbation. The vector $\mathbf{r}_j$ depends on both $\lambda_j$ and time $t$ through the varying orbital elements of planet~$j$.
   
   The heliocentric Keplerian elements of the small body are written $(a,e,i,\omega,\Omega,M)$, with the associated canonical Delaunay elements:
   \begin{equation}
      \left\{
      \begin{aligned}
         \ell &= M \,,\\
         g &= \omega \,,\\
         h &= \Omega\,,
      \end{aligned}
      \right.
      \quad\quad
      \left\{
      \begin{aligned}
         L &= \sqrt{\mu a}\,, \\
         G &= \sqrt{\mu a(1-e^2)}\,, \\
         H &= \sqrt{\mu a(1-e^2)}\cos i\,.
      \end{aligned}
      \right.
   \end{equation}
   We consider a mean-motion resonance $k_p$:$k$ between the small body and a given planet with index $j=p$, where $k_p$ and $k$ are co-prime integers. At the vicinity of the resonance, the combination $\sigma = k\lambda-k_p\lambda_p$ varies slowly, where $\lambda=\ell+g+h$. The combination $\sigma$ is taken as a new canonical coordinate by a linear unimodular transformation \citep{Milani-Baccili_1998}. We use the following choice of coordinates:
   \begin{equation}\label{eq:rescoo1}
      \begin{pmatrix}
         \sigma \\
         \gamma \\
         u \\
         v
      \end{pmatrix}
      =
      \begin{pmatrix}
         k & -k_p & k & k \\
         c & -c_p & c & c \\
         0 &    0 & 1 & 0 \\
         0 &    0 & 0 & 1
      \end{pmatrix}
      \begin{pmatrix}
         \ell \\
         \lambda_p \\
         g \\
         h
      \end{pmatrix}\,,
   \end{equation}
   which leads to the new momenta
   \begin{equation}\label{eq:rescoo2}
      \begin{pmatrix}
         \Sigma \\
         \Gamma \\
         U \\
         V
      \end{pmatrix}
      =
      \begin{pmatrix}
           -c_p & -c & 0 & 0 \\
            k_p &  k & 0 & 0 \\
            -1  &  0 & 1 & 0 \\
            -1  &  0 & 0 & 1
      \end{pmatrix}
      \begin{pmatrix}
         L \\
         \Lambda_p \\
         G \\
         H
      \end{pmatrix}\,.
   \end{equation}
   In this expression, $c$ and $c_p$ are integers chosen such that $c\,k_p-c_p\,k = 1$; they always exist as $\gcd(k,k_p)=1$.
   
   This specific choice for the variable $\sigma$ emphasises that we do not presuppose any particular form for the resonance angle, and keep all resonant terms with combination $k_p$:$k$ in the Hamiltonian function. If we assume that the small body is close to or inside a resonance with combination $k_p$:$k$, then the new coordinates $\gamma$ and $\{\lambda_{j\neq p}\}$ are fast angles (orbital timescale; frequency $\propto \varepsilon^0$), $\sigma$ is a semi-slow angle (resonant timescale; frequency $\propto\varepsilon^{1/2}$), and $(u,v)$ are slow angles (secular timescale; frequency $\propto\varepsilon^1$; see Sect.~\ref{asec:adiab} below).
   
   \subsection{The semi-secular system}\label{sec:semisec}
   We define the semi-secular Hamiltonian function $\mathcal{K}$ by removing the fast angles from the Hamiltonian $\mathcal{H}$ in Eq.~\eqref{eq:Hfirst} using a close-to-identity change of coordinates. Therefore, in the semi-secular coordinates (that we write with the same symbols for simplicity), the Hamiltonian function does not depend on $\gamma$ and $\{\lambda_{j\neq p}\}$. As a consequence, the momenta $\Gamma$ and $\{\Lambda_{j\neq p}\}$ are constants of motion which have arbitrary values. We conveniently choose them equal to zero, such that our coordinates are simply
   \begin{equation}\label{eq:SUV}
      \left\{
      \begin{aligned}
         \sigma &= k\lambda - k_p\lambda_p\,,\\
         u &= \omega\,, \\
         v &= \Omega\,,
      \end{aligned}
      \right.
      \quad\quad
      \left\{
      \begin{aligned}
         \Sigma &= \frac{\sqrt{\mu a}}{k}\,, \\
         U &= \sqrt{\mu a}\left(\sqrt{1-e^2}-1\right)\,, \\
         V &=  \sqrt{\mu a}\left(\sqrt{1-e^2}\cos i-1\right)\,.
      \end{aligned}
      \right.
   \end{equation}
   Dropping constant terms, the semi-secular Hamiltonian function is defined as
   \begin{equation}\label{eq:K}
      \mathcal{K} = \mathcal{K}_0(\Sigma) + \varepsilon\mathcal{K}_1(\Sigma,U,V,\sigma,u,v) + \mathcal{O}(\varepsilon^2)\,,
   \end{equation}
   where
   \begin{equation}\label{eq:Kdet}
      \left\{
      \begin{aligned}
         \mathcal{K}_0 &= -\frac{\mu^2}{2(k\Sigma)^2} - n_pk_p\Sigma\,, \\
         \mathcal{K}_1 &= - \sum_{j\neq p}\frac{\mu_j}{4\pi^2}\int_{0}^{2\pi}\!\!\!\int_{0}^{2\pi}\frac{1}{\|\mathbf{r}-\mathbf{r}_j\|}\,\mathrm{d}\ell\,\mathrm{d}\lambda_j \\
         &\ \ \ - \frac{\mu_p}{2\pi}\int_{0}^{2\pi}\left(\frac{1}{\|\mathbf{r}-\mathbf{r}_p\|}-\mathbf{r}\cdot\frac{\mathbf{r}_p}{\|\mathbf{r}_p\|^3}\right)\mathrm{d}\gamma\,.
      \end{aligned}
      \right.
   \end{equation}
   For the planets with index $j\neq p$, the indirect part vanishes over the average so we omit it in Eq.~\eqref{eq:Kdet}. Apart from $\lambda_j$, the orbital elements of planet~$j$ are slowly-varying coordinates that we keep constant while computing the integrals. We compute the integrals numerically, so that all resonant terms are included and our approach is valid for all eccentricities and inclinations.
   
   Upon averaging, the orbital elements and mean-motions of the planets used in Eq.~\eqref{eq:Kdet} become secular variables that incorporate the effects of mutual perturbations among planets.
   
   \subsection{Expansion around the resonance centre}
   Near the $k_p$:$k$ resonance with planet $p$, the conjugate pair $(\Sigma,\sigma)$ evolves with characteristic frequency or order $\varepsilon^{1/2}$, while the secular precession of the orbits takes place with frequency of order $\varepsilon$. Hence, if $\varepsilon$ is small enough, these two timescales are well separated and we can use the same averaging method as in Sect.~\ref{sec:semisec} to remove the angle $\sigma$ from the Hamiltonian function. However, the two characteristic frequencies are only separated by a factor $\varepsilon^{1/2}$ so the neglected terms are of order $\varepsilon^{3/2}$ (and not $\varepsilon^2$ as in Eq.~\ref{eq:K}).
   
   In practice, this method consists in first studying the evolution of $(\Sigma,\sigma)$ for fixed values of $(U,V,u,v)$ and the planets' mean orbital elements. It is called the `adiabatic approximation' and was popularised notably by \cite{Wisdom_1985} and \cite{Henrard_1982,Henrard_1993}. Here, we only need to compute the resonance widths, which amounts to finding numerically the separatrices of the resonance in the $(\Sigma,\sigma)$ plane.
   
   In order to find the resonance separatrices, it is convenient to first expand the Hamiltonian function at the vicinity of the resonance centre\footnote{This step is not compulsory; one can also directly study the Hamiltonian function $\mathcal{K}$ in Eq.~\eqref{eq:Kdet} in the two-dimensional plane spanned by $(\Sigma,\sigma)$ as done by \cite{Saillenfest-etal_2016} and \cite{Lei-etal_2022}. However, this introduces an unnecessary complexity in the numerical algorithm.}. In the restricted three-body problem, we usually estimate the resonance centre $\Sigma_0$ from the Hamiltonian $\mathcal{K}_0$ taken alone. However, in a multi-planetary system as considered here, there can be a constant shift between this value and the one obtained by considering the whole Hamiltonian. This constant shift is due to the $0$th-order term of the development of $\mathcal{K}_1$ in Legendre polynomials. It can be taken into account by redefining the splitting of the Hamiltonian:
   \begin{equation}
      \left\{
      \begin{aligned}
         \mathcal{K}_0' &= \mathcal{K}_0(\Sigma) -\frac{\mu}{(k\Sigma)^2}\sum_{j\in\mathcal{I}}\varepsilon\mu_j\,, \\
         \mathcal{K}_1' &= \mathcal{K}_1(\Sigma,U,V,\sigma,u,v) + \frac{\mu}{(k\Sigma)^2}\sum_{j\in\mathcal{I}}\mu_j\,,
      \end{aligned}
      \right.
   \end{equation}
   where the additional term is the $0$th-order term in the Legendre development of $\mathcal{K}_1$, that we transfer to $\mathcal{K}_0$ (and thus remove from $\mathcal{K}_1$). Here, the summation is only realised over the subset $\mathcal{I}$ of planets inner to the small body, since the outer planets have no $0$th-order dependence on the semi-major axis $a$ (see e.g. \citealp{Laskar-Boue_2010}). As defined from $\mathcal{K}_0'$ taken alone, the nominal semi-major axis at the centre of the resonance is
   \begin{equation}
      \frac{\partial\mathcal{K}_0'}{\partial\Sigma}(\Sigma_0) = 0 \iff \Sigma_0^3 = \frac{\mu^2}{n_pk_pk^2}\left(1 + 2\varepsilon\sum_{j\in\mathcal{I}}\frac{\mu_j}{\mu}\right)\,.
   \end{equation}
   From Eq.~\eqref{eq:SUV}, $\Sigma_0$ is equivalent to a semi-major axis $a_0$ as expressed in Eq.~\eqref{eq:a0} of the main text. If we set $\varepsilon=0$, we retrieve the classical estimate of the resonance centre. We now expand the Hamiltonian function around $\Sigma_0$:
   \begin{equation}
      \begin{aligned}
         \mathcal{K} &= \mathcal{K}_0'(\Sigma_0) + \frac{1}{2}(\Sigma-\Sigma_0)^2\frac{\partial^2\mathcal{K}_0'}{\partial\Sigma^2}(\Sigma_0) + \mathcal{O}\left((\Sigma-\Sigma_0)^3\right)\\
         & + \varepsilon\mathcal{K}_1'(\Sigma_0,U,V,\sigma,u,v) + \mathcal{O}\Big(\varepsilon(\Sigma-\Sigma_0)\Big)\,.
      \end{aligned}
   \end{equation}
   Considering that the small body is close to or inside the resonance, the distance $|\Sigma-\Sigma_0|$ is of the order of the resonance width\footnote{This behaviour of the resonance width is not true for near-zero eccentricities in first-order resonances. The shift of the resonance centre and opening of the separatrices when $e\rightarrow 0$ (see e.g. \citealp{Henrard-Lemaitre_1983,Murray-Dermott_1999,Malhotra-Zhang_2020,Malhotra-Chen_2023}) are therefore not captured in the model presented here.} which goes as $\varepsilon^{1/2}$ (this well known property can be verified a posteriori; see Eq.~\ref{eq:width} of the main text). As we neglect terms of order $\varepsilon^{3/2}$ when using the adiabatic approximation, it is unnecessary to keep them at this point. Therefore, omitting constant parts, we obtain
   \begin{equation}\label{eq:Ktrick}
      \mathcal{K} = -\frac{1}{2}\alpha(\Sigma-\Sigma_0)^2 + \varepsilon W(U,V,u,v,\sigma) + \mathcal{O}\left(\varepsilon^{3/2}\right)\,,
   \end{equation}
   where $\alpha=\mathcal{O}(1)$ is a positive constant expressed in Eq.~\eqref{eq:alp} of the main text, and
   \begin{equation}\label{eq:W}
      W(U,V,u,v,\sigma)=\mathcal{K}_1'(\Sigma_0,U,V,\sigma,u,v)\,.
   \end{equation}
   The Hamiltonian function in Eq.~\eqref{eq:Ktrick} is given in a slightly different form in Eqs.~\eqref{eq:KW} and \eqref{eq:WC} of the main text.
   
   \subsection{The adiabatic approximation}\label{asec:adiab}
   Writing down Hamilton's equations of motion and using the variable $\Phi = (\Sigma-\Sigma_0)/\sqrt{\varepsilon}=\mathcal{O}(1)$ instead of $\Sigma$, we get
   \begin{equation}\label{eq:timescale}
      \begin{aligned}
         \dot{\sigma} &= -\sqrt{\varepsilon}\,\alpha\,\Phi\,, \\
         \dot{\Phi} &= -\sqrt{\varepsilon}\,\frac{\partial W}{\partial\sigma}\,,
      \end{aligned}
      \quad\quad
      \begin{aligned}
         \dot{u} &= \varepsilon\,\frac{\partial W}{\partial U}\,, \\
         \dot{U} &= -\varepsilon\,\frac{\partial W}{\partial u}\,,
      \end{aligned}
      \quad\quad
      \begin{aligned}
         \dot{v} &= \varepsilon\,\frac{\partial W}{\partial V}\,, \\
         \dot{V} &= -\varepsilon\,\frac{\partial W}{\partial v}\,.
      \end{aligned}
   \end{equation}
   As pointed out by \cite{Sidorenko_2006}, the pair $(\Phi,\sigma)$ may be used as conjugate coordinates if we divide the variables $(U,V,u,v)$ by $\varepsilon^{1/4}$. In Eq.~\eqref{eq:timescale}, the difference of timescale between the three degrees of freedom clearly appears: $(\Phi,\sigma)$ are semi-slow (time-derivatives of order $\sqrt{\varepsilon}$) whereas $(U,V,u,v)$ are slow (time derivatives of order $\varepsilon$). This leads us to the adiabatic approximation, that is, solving for the semi-slow degree of freedom $(\Phi,\sigma)$ while fixing the slow ones $(U,u)$ and $(V,v)$.
   
   In practice, the coordinates $U$ and $V$ defined in Eq.~\eqref{eq:SUV} can be replaced by equivalent variables $\bar{e}$ and $\bar{i}$ as
   \begin{equation}
      \begin{aligned}
         U &= \sqrt{\mu a_0}\left(\sqrt{1-\bar{e}^2} - 1\right)\,,\\
         V &= \sqrt{\mu a_0}\left(\sqrt{1-\bar{e}^2}\cos\bar{i} - 1\right)\,.\\
      \end{aligned}
   \end{equation}
   For a small body oscillating inside the resonance, $\bar{e}$ and $\bar{i}$ can be directly interpreted as the mean eccentricity and inclination of the small body over one libration cycle. For simplicity, we drop the `bar' symbol in the main text and identify $e$ and $i$ as $\bar{e}$ and $\bar{i}$.
   
   As $\mathcal{K}_1'$ is evaluated at $a=a_0$ in the definition of $W$ (see Eq.~\ref{eq:W}), we can now perform the required integrals in Eq.~\eqref{eq:Kdet} using the standard Keplerian elements $(a_0,e,i,\omega,\Omega)$ and forget about the canonical coordinates $(U,V,u,v)$ taken as constants. The Hamiltonian function in Eq.~\eqref{eq:Ktrick} can then be used to compute the characteristics of the $k_p$:$k$ mean-motion resonance, and in particular its widths, as described in Sect.~\ref{sec:methwidth} of the main text.